\documentclass[prl,aps,twocolumn,amsmath,amssymb,superscriptaddress,showpacs,showkeys]{revtex4-1}
\usepackage{graphicx}
\usepackage{dcolumn}
\usepackage{bm}
\usepackage{float}
\usepackage[colorlinks=true, allcolors=blue, linktocpage=true]{hyperref}

\begin{document}

\title{Near-unity indistinguishability single photon source for large-scale\\ integrated quantum optics}

\author{\L{}ukasz Dusanowski}
\email{lukasz.dusanowski@physik.uni-wuerzburg.de}
\affiliation{Technische Physik, University of W\"{u}rzburg, Physikalisches Institut and Wilhelm-Conrad-R\"{o}ntgen-Research Center for Complex Material Systems, Am Hubland, D-97074 W\"{u}rzburg, Germany}

\author{Soon-Hong Kwon}
\affiliation{Technische Physik, University of W\"{u}rzburg, Physikalisches Institut and Wilhelm-Conrad-R\"{o}ntgen-Research Center for Complex Material Systems, Am Hubland, D-97074 W\"{u}rzburg, Germany}
\affiliation{Department of Physics, Chung-Ang University, 156-756 Seoul, Korea}

\author{Christian Schneider}
\affiliation{Technische Physik, University of W\"{u}rzburg, Physikalisches Institut and Wilhelm-Conrad-R\"{o}ntgen-Research Center for Complex Material Systems, Am Hubland, D-97074 W\"{u}rzburg, Germany}

\author{Sven H\"{o}fling}
\affiliation{Technische Physik, University of W\"{u}rzburg, Physikalisches Institut and Wilhelm-Conrad-R\"{o}ntgen-Research Center for Complex Material Systems, Am Hubland, D-97074 W\"{u}rzburg, Germany}
\affiliation{SUPA, School of Physics and Astronomy, University of St Andrews, KY16 9SS St Andrews, UK}

\date{\today}


	
\begin{abstract}
Integrated single photon sources are key building blocks for realizing scalable devices for quantum information processing. For such applications highly coherent and indistinguishable single photons on a chip are required. Here we report on a triggered resonance fluorescence single photon source based on In(Ga)As/GaAs quantum dots coupled to single- and multi-mode ridge waveguides. We demonstrate the generation of highly linearly polarized resonance fluorescence photons with 99.1\% (96.0\%) single-photon purity and 97.5\% (95.0\%) indistinguishability in case of multi-mode (single-mode) waveguide devices fulfilling the strict requirements imposed by multi-interferometric quantum optics applications. Our integrated triggered single photon source can be readily scaled up, promising a realistic pathway for on-chip linear optical quantum simulation, quantum computation and quantum networks.  
\end{abstract}

\maketitle

Large scale implementations of quantum information processing (QIP) schemes are one of the major challenges of modern quantum physics. Building a platform successfully combing many qubits is a very demanding task, however it would provide a system capable to perform quantum simulations, quantum computing and secure quantum communication. In this regard, using single photons as qubits is a particularly appealing concept. Due to the photons low decoherence and the inherent possibility of low-loss transmission, they can be used for both quantum computing and quantum communication applications~\cite{kok2010introduction}. Over the last decade, there have been extensive experimental efforts towards realizing large-scale optical QIP systems. A vast majority of these implementations, such as linear optics quantum computing~\cite{Knill2001}, boson sampling~\cite{Spring2013} or quantum repeater schemes~\cite{Briegel1998} involve a two-photon interference effect, where a complete wave-packet overlap of single photons at the beam splitter is required. This feature demands photons, which are indistinguishable in terms of energy, bandwidth, polarization and arrival time at the beam splitter. Consequently, sources of indistinguishable single photons are one of the central resources for a large scale experimental realization of the optical QIP devices.

Among different kinds of emitters quantum dots (QDs) coupled to photonic structures have been shown to be one of the brightest single photon sources (SPS)~\cite{Santori2002,Aharonovich2016,Senellart2017}, which under resonant excitation conditions~\cite{Flagg2009,Ates2009,Senellart2017} can reach simultaneously indistinguishabilities higher than 95\%, single photon purities better than 99\% and extraction efficiencies as high as 65-79\%~\cite{Ding2016,Somaschi2016,Unsleber2016,Senellart2017,Fischer2016}. Further, by applying advanced semiconductor micro-processing technologies it is possible to fabricate devices where QD electronic properties can be dynamically shaped by strain~\cite{Zhang2016,Trotta2016} or electric~\cite{Kirsanske2017b,Hoang2012} fields. The optical quality of such sources allowed already for demonstration of on-demand CNOT-gates~\cite{Pooley2012,Gazzano2013a,He2013a}, heralded entanglement between distant hole spins~\cite{Delteil2015} or the recent realization of 3-,4- and 5-photon boson sampling~\cite{Wang2017,Loredo2017}. 

It is believed that future steps towards large scale quantum optics should ensure the full on-chip scalability of SPSs~\cite{Dietrich2016a,Aharonovich2016}. A natural system towards this goal are integrated circuits, where QD SPSs can be homogeneously~\cite{Jons2015,Enderlin2012,Schwagmann2011,Arcari2014,Reithmaier2015} or heterogeneously~\cite{Davanco2017,Elshaari2017,Kim2017} integrated on a single chip. In this approach light can be directly coupled into in-plane waveguides (WGs) and combined with other functionalities on a chip such as phase shifters~\cite{Midolo2017,Wang2014a}, beam splitters~\cite{Prtljaga2014,Jons2015,Kim2017}, filters~\cite{Elshaari2017,Harris2014}, detectors~\cite{Reithmaier2015,Kaniber2016} and other devices for light propagation, manipulation and detection on a single photon level. 

\begin{figure}
	\includegraphics[width=\columnwidth]{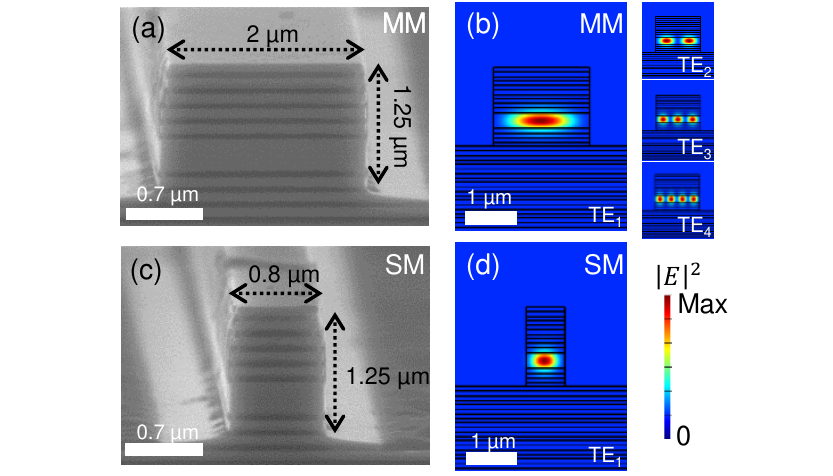}
	\caption{\label{fig:1} Ridge waveguide image and mode profile. Scanning electron microscope images of the processed (a)~multi-mode (MM) and (c)~single-mode (SM) ridge waveguide structure. Optical power distribution profile for the TE (b)~fundamental - TE$_1$ and higher order - TE$_{2-4}$ modes of MM and (d)~fundamental TE$_1$ mode of SM waveguide at 930~nm wavelength.}
\end{figure}

By utilizing this idea near-unity coupling efficiency of a QD emitter to waveguide device was already achieved~\cite{Lund-Hansen2008,Enderlin2012,Arcari2014,Stepanov2015}, showing the undeniable potential of this concept. In addition, integrated circuits allow to spatially separate excitation and detection spots, which straightforwardly enables applying resonant driving schemes to slow-down decoherence processes and reduce on-demand emission time-jitter. This technique was already applied to waveguide integrated QDs under both continuous-wave~\cite{Makhonin2014,Kalliakos2016} (CW) and pulsed~\cite{Schwartz2016,Kirsanske2017b,Liu2018} excitation. In particular, two-photon interference of subsequently emitted resonance fluorescence photons have been demonstrated recently~\cite{Kirsanske2017b,Liu2018}, however recorded indistinguishability values failed to reach requirements imposed by quantum optics applications~\cite{Pan2012}. 

One of the major issues in QD-based on-chip SPSs is the loss of photon indistinguishability related to the charge fluctuation from nearby etched surfaces~\cite{Makhonin2014, Kalliakos2016}. This is true especially for small in size, complex structures such as photonic crystal~\cite{Liu2018} or nanobeam~\cite{Kirsanske2017b} waveguides. To diminish this effect a number of strategies can be employed. Firstly, the amount of surface states could be decreased by optimizing passivation of the etched surfaces~\cite{Press2010}. Secondly, the charge environment could be stabilized by weak CW non-resonant optical illumination~\cite{Majumdar2011} or gating~\cite{Somaschi2016, Liu2018}. Finally, the Purcell effect might be used to enhance the radiative emission rate and thus improve the photon indistinguishability in the presence of dephasing~\cite{Somaschi2016,Ding2016,Unsleber2016,Liu2018,Iles-Smith2017}. As we will show in this Letter, a simplified waveguide design with relatively large profile dimensions, keeping etched surfaces far away from QD, can be also very advantageous in this respect. By utilizing distributed Bragg reflectors (DBR) ridge waveguide design we fabricated SPSs which simultaneously meet the requirements of near perfect single photon purity and indistinguishability.            

To evaluate the performance of our devices we performed resonance fluorescence experiments on In(Ga)As/GaAs QDs coupled to single-mode (SM) and multi-mode (MM) in-plane waveguides. The light confinement and guiding were achieved by DBRs in vertical and ridges defined in horizontal direction. Usage of DBRs WG instead of the GaAs WG slab approach allowed us to soften the waveguide profile dimensions allowing to achieve single-mode operation while keeping a relatively high QD-waveguide light coupling efficiency (14-19\% into one WG arm). To simulate the integrated circuit device operation, the QDs were excited resonantly from top of the waveguides, and the emitted photons were collected from the side facets after up to 1~mm travel distance on a chip. Under such conditions a significant reduction of the scattered laser intensity was achieved, enabling our MM (SM) waveguide device generation of record high on-chip 97.5\% (95.0\%) indistinguishable triggered single photons with a 99.1\% (96.0\%) single-photon purity and over 99\% (98\%) linear polarization. We believe, that this integrated SPSs can be readily scaled up demonstrating a realistic pathway for on-chip optical quantum processing.

To realize our waveguide integrated SPS we have grown In(Ga)As/GaAs QDs embedded in a low-quality factor cavity (Q~$\sim$~200) based on DBRs. By performing three-dimensional finite-difference-time-domain (FDTD) calculations we investigated different waveguide designs for maximized coupling efficiency. We found that 1.25-1.65~$\mu$m WG height and 0.6-2.0~$\mu$m width lay within optimal values and allow to achieve around 10-22\% coupling efficiency into each WG arm (total 20-44\%). Moreover, our DBR waveguides can be operated in the single-mode regime for WG widths as small as 0.9~$\mu$m at 900~nm cut-off wavelengths. Based on those considerations, we realized two types of ridge waveguides: (i)~MM WGs with 2.0~x~1.25~$\mu$m$^2$ profile and (ii)~SM WGs with 0.8~x~1.25~$\mu$m$^2$, for which we expect $\sim$14\% and $\sim$19\% QD coupling efficiency into one WG arm, respectively. We point out, that in principle the QD-emission coupling efficiency could be further improved by integration of DBR WGs with low-refractive-index layers while maintaining the mentioned relatively large size of WG profile. More details can be found in Supplemental Materials. Scanning electron microscope images of our fabricated MM and SM ridge waveguides are shown in Figures~\ref{fig:1}(a) and (c), respectively. Figures~\ref{fig:1}(b) and (d) show simulated optical mode profiles of the light field confined in our devices, calculated for the transverse-electric (TE) modes at 930~nm. In both cases the mode profiles are confined within the defined ridges, allowing for the single photon guiding along the chip. The modes are mainly concentrated in the GaAs cores and partially penetrate the top and bottom DBR mirrors.”

\begin{figure}
	\includegraphics[width=\columnwidth]{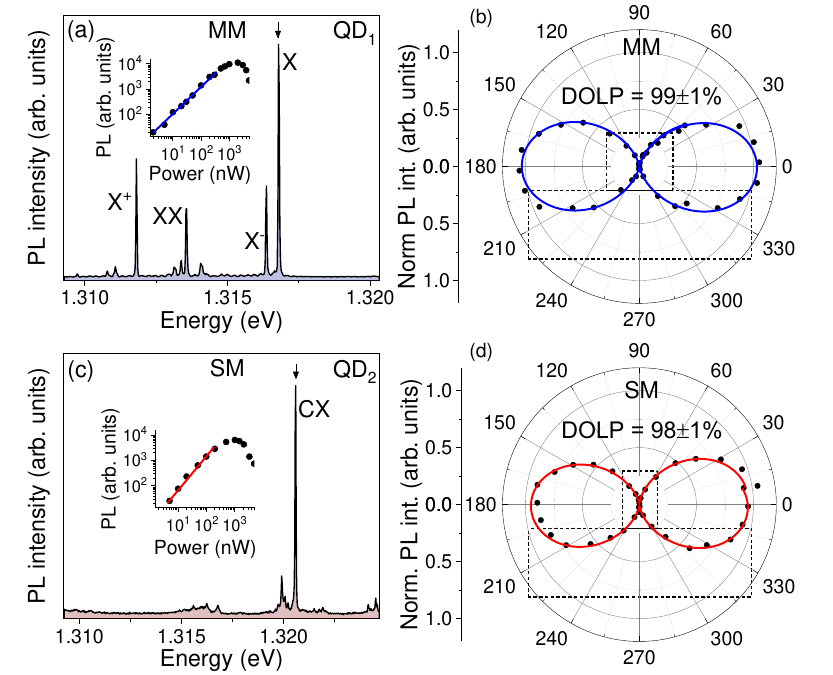}
	\caption{\label{fig:2} Non-resonant photoluminescence. Side collected QD emission spectra recorded from (a) MM and (c) SM ridge WG device at 1~$\mu$W power CW excitation. Insets: intensity vs power dependencies of the marked with arrows PL peaks in log-log scale. Red/blue solid curve: fit with a power function showing linear dependence. Polarization characteristics of the (b)~MM and (d)~SM WG coupled QD PL emission, revealing 99$\pm$1\% and 98$\pm$1\% degree of linear polarization, respectively, oriented along the TE mode of the ridge waveguides.}
\end{figure}

Initially, both devices were characterized optically under non-resonant excitation conditions. Figure~\ref{fig:2}(a) and (c) show side collected photoluminescence (PL) spectra from a QD$_1$ and QD$_2$, respectively, under above-bandgap CW pump (660~nm diode laser) from the top of the waveguide. In case of QD$_1$ coupled to a MM WG four intense emission lines are visible, where the one of interest centered at 1.3169~eV (marked with an arrow) was identified as a neutral exciton (X). The inset in Fig.~\ref{fig:2}(a) shows central peak intensity changes vs excitation power indicating a clear linear dependence. The remaining emission lines have been identified based on power- and top-detected-polarization-resolved PL as positively charged exciton (X$^+$), negatively charged exciton (X$^-$) and biexciton (XX) recombination from the same QD. Spectra for QD$_2$ coupled to the SM WG consists of a single emission line centered at 1.3206~eV, identified as a charged exciton (CX). Both studied emission lines show a high degree of linear polarization (DOLP) of around 99$\pm$1\% and 98$\pm$1\% for QD$_1$ and QD$_2$, respectively, oriented in sample plane as shown in Figures~\ref{fig:2}(b) and (d). A high DOLP and its direction are related to the QDs dipole moments, which are mainly in-plane oriented and thus emitted photons mostly couple to and propagate in the TE waveguide mode.  

\begin{figure}[b]
	\includegraphics[width=\columnwidth]{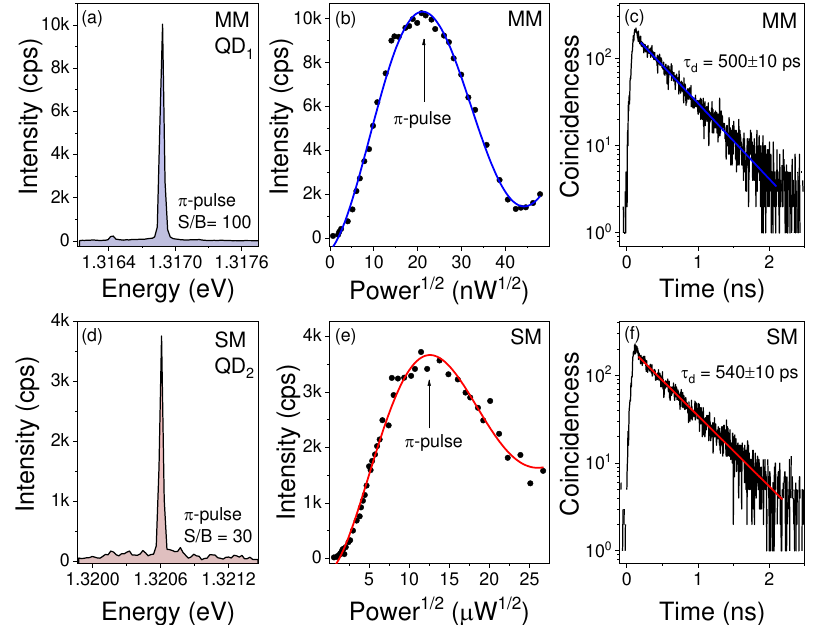}
	\caption{\label{fig:3} Pulsed resonance fluorescence. Side collected QD emission spectra from (a) MM and (d) SM waveguide device under $\pi$-pulse excitation. (b),(e)~Signal intensity versus square root of incident power. Solid red/blue curve: fit with a dumped sinusoidal function. (c),(f)~Time-resolved resonance fluorescence measurement under $\pi$-pulse pumping. Red/blue solid curves: fit using a mono-exponential decay function with time constants of 500$\pm$10~ps and 540$\pm$10~ps for MM and SM device, respectively.}
\end{figure}

Next, the characteristics of the devices were probed under pulsed s-shell resonant excitation. Figures~\ref{fig:3}(a) and (d) show side detected pulsed resonance fluorescence spectra displayed on a spectrometer at a temperature of 4.5~K. Since in our experimental configuration the excitation and collection spots are spatially separated by hundreds of micrometers, it allows us to rather readily suppress the stray laser photons scattered from the excitation area. This is usually not the case for short length waveguides as the laser light scattered from the excitation area might be collected within the numerical aperture of the detection objective. Additionally, the presence of low Q factor cavity in the plane of the sample allowed us to more effectively excite the emitters and thus reduce powers needed for resonant driving. To further suppress the influence of laser light scattering on the single photon performance, the beam spot size, as well as polarization, was carefully controlled. By utilizing both the intrinsic (spatial separation) and polarization filtering we were able to obtain a signal-to-background (S/B) ratio of over 100 for MM and 30 for SM waveguide devices under $\pi$-pulse excitation. In fact, we believe that polarization filtering can be omitted for fully on-chip device operation.  

In Figures~\ref{fig:3}(b) and (e) the resonance fluorescence intensity versus the square root of the incident power are shown. Clear Rabi oscillations with visible damping for both QDs are observed, which is due to coherent control of the particular QD's two-level systems coupled to phonon bath~\cite{Forstner2003}. The emission intensities in both cases reach the maximum for $\pi$-pulse with laser powers of 440~nW and 160~$\mu$W for QD$_1$ and QD$_2$, respectively. The significantly larger pump power required to reach $\pi$-pulse for QD$_2$ is most likely related to the smaller size of the waveguide in respect to the laser beam spot size, as well as a slight energy detuning from the planar cavity resonance. The resonance fluorescence intensity of around 10~kcps (3.5~kcps) was observed on the avalanche photodiode detector (setup efficiency~$\sim$2\%) at $\pi$-pulse for a MM (SM) device, which corresponds to $\sim$0.6\% ($\sim$0.2\%) total photon extraction efficiency from a QD collected by the first lens, and $\sim$12\% ($\sim$2\%) coupling efficiency into one WG arm (lower bound estimated based on 95\% and 90\% losses due to the out-coupling). It needs to be noted that the design of the WGs for the high out-coupling into external collection optics was not optimized since ultimately all single photon processing is supposed to be performed on-chip. Under $\pi$-pulse excitation the time-resolved resonance fluorescence measurements have been performed. The recorded fluorescence decay time traces shown in Fig.~\ref{fig:3}(c) and (f) demonstrate clear mono-exponential decays with the time constants of 500$\pm$10~ps and 540$\pm$10~ps for QD$_1$ and QD$_2$, respectively. 
          
\begin{figure}
	\includegraphics[width=\columnwidth]{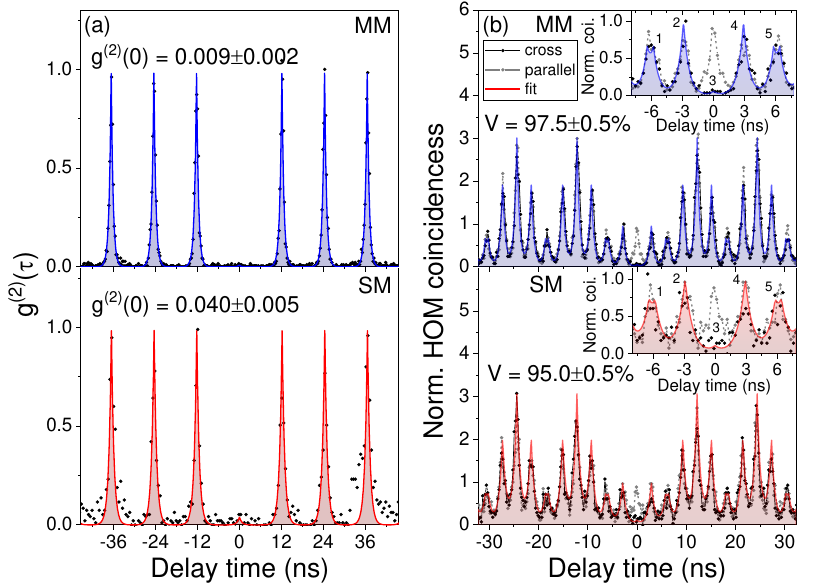}
	\caption{\label{fig:4} Single photon generation and two photon interference under resonant $\pi$-pulse excitation. (a)~Side collected resonance fluorescence intensity-correlation histogram recorded for QD$_1$ coupled to MM WG (upper panel) and QD$_2$ coupled to SM WG (lower panel). $g^{(2)}(0)$ values are calculated from the integrated photon counts, while the uncertainty is based on the standard deviation of the Poissonian peak counts. (b)~Two-photon interference HOM histogram recorded for the 3~ns time separated co- (black points) and cross-polarized (grey points) single photons recorded for QD$_1$ (upper panel) and QD$_2$ (lower panel). Red/blue solid curves: fits based on two-sided exponential decay functions.}
\end{figure}

In order to characterize purity and indistinguishability of our SPSs auto-correlation and two-photon interference experiments have been performed on the resonance fluorescence signal filtered out from a broader laser profile and phonon sidebands. In Fig.~\ref{fig:4}(a) second-order correlation function histograms recorded in Hanbury Brown and Twiss (HBT) configuration under $\pi$-pulse excitation for QD$_1$ and QD$_2$ are shown. In both cases, nearly vanishing multi-photon emission probabilities at zero delays have been recorded with $g^{(2)}(0)=0.009\pm0.002$ and $g^{(2)}(0)=0.04\pm0.005$ for QD$_1$ and QD$_2$, respectively. The shape of all the peaks exhibits a clear two-sided mono-exponential decay with a time constant corresponding to the decay time recorded directly in the time-resolved resonance fluorescence measurements.

To study the indistinguishability of the emitted photons the QDs were excited twice every repetition cycle (12.2~ns) by a pair of pulses separated by 3~ns. Two subsequently emitted photons are then introduced into a 3~ns unbalanced interferometer where a delay between them is compensated in order to superimpose single photon pulses on the beam splitter~\cite{Santori2002}. If the two photons are perfectly indistinguishable they will always exit the same but a random output port, which is quantitatively translated into the two-coincidences correlation dip at a zero delay. Hong-Ou-Mandel (HOM) correlation histograms obtained for the two considered photon sources are presented in Figure~\ref{fig:4}(b) in upper (QD$_1$) and lower (QD$_2$) panels. The histograms consist of a set of five 3~ns delayed peak clusters separated by the repetition time of the laser (first and the last peak of the neighboring clusters are superimposed). The central cluster [insets in Figure~\ref{fig:4}(b)] describes coincidence events related to the single-photons traveling through different paths of the interferometer. This is described in detail in the Supplemental Materials. In order to evaluate the zero delay peak (no.~3) area with respect to the neighboring peaks (no.~2 and 4) the experimental data have been fitted with the two-side exponential decay functions. Upon this procedure, the two-photon HOM interference visibility of 0.975$\pm$0.005 for QD$_1$ and 0.950$\pm$0.005 for QD$_2$ have been obtained after correcting for HOM setup imperfections such as beam-splitting ratio (R/T~=~1.15) and contrast of the Mach-Zehnder-interferometer (1-$\varepsilon$~=~0.99)\cite{Santori2002,Fischer2016}.  

It was recently demonstrated, that single photon emission purity and indistinguishability in resonantly excited two-level systems is intrinsically limited by the re-excitation process\cite{Hanschke2018,Fischer2016}. Specifically, it was shown that the laser excitation pulse-length $\tau_{pulse}$ sets the lower bounds of the HBT and HOM experimental two-photon coincidences probability, and thus $g^{(2)}(0)$ and visibility values obtainable for a given SPS with characteristic emission time $\tau_{emitter}$. Based on reference \cite{Fischer2016} those bounds can be calculated following the linear dependence $0.4\cdot\tau_{pulse}/\tau_{emitter}$, which in our case limits $g^{(2)}(0)$ (visibility) to 0.0016 (0.9984) and 0.0015 (0.9985) for QD$_1$ and QD$_2$, respectively. Since the experimentally obtained single-photon purity values are significantly above the aforementioned limits, we believe their dominant component might be non-filtered residual scatter of the laser pulse from the sample surface, which we do not take into account in HBT and HOM data analysis.

Imperfect efficiency and indistinguishability of SPSs are related to the problem of optical quantum computation under a degree of experimental error. There have been a number of promising proposals of linear optical quantum computing that are robust against imperfect SPSs and inefficient detectors. In particular, it was shown, that fault-tolerant quantum computation can be performed if the two-photon gate operation error probability is lower than a 1\% threshold value~\cite{Fowler2009,Knill2005}. In this regard, we can calculate how the indistinguishability of input photons affects the two-photon gate performance assuming perfect beam-overlap, alignment, beam-splitters and no dark counts in single photon counting detection. In case of our MM (SM) waveguide QD source with 97.5\% (95.0\%)  photons visibility a gate fidelity of 99.5\% (99.1\%)~\cite{Wei2014a} is theoretically obtainable. Those values already surpass mentioned 1\% precision threshold value, and in this context, our work provides SPSs with photons visibility needed for scalable quantum technologies.

Another source of error in quantum optics, which is far more dominant than gate fidelity is photon loss. This problem is associated with the overall source and detectors efficiency, which product has to be greater than 2/3 in order to perform efficient fault-tolerant linear optical quantum computation~\cite{Varnava2008}. In principle, this efficiency threshold is very difficult to fulfill in any system, since every optical setup exhibit losses. In all on-chip platforms however, where single photons are generated, routed, manipulated and eventually detected within the same low loss photonic circuit, this efficiency threshold is likely to be fulfilled within the near future.   

In this context, essential next steps to make QD-based on-chip platform feasible for fully scalable quantum technologies would consist of (i)~improving QD-waveguide circuit coupling efficiency while maintaining high degree of photons indistinguishability, (ii)~introducing high-visibility and low-loss on-chip interferometers and phase shifters based on single-mode waveguides, and (iii)~integrating with high efficiency superconducting detectors~\cite{Dietrich2016a,Aharonovich2016,Senellart2017}. Such a system may at some point overcome the intrinsic limitations of the vertical devices, opening the possibility to create a scalable quantum integrated circuits operating at the single photon level.

In this Letter, we have shown that our MM (SM) waveguide integrated SPS can generate photons with near-unity indistinguishability of 0.975$\pm$0.005 (0.950$\pm$0.005) along with the $g^{(2)}(0)$ value equal to 0.009$\pm$0.002 (0.040$\pm$0.005). We demonstrated single photon propagation of over hundreds of micrometers in waveguides and QD-WG coupling efficiency of $\sim$12\% ($\sim$2\%) into one WG arm. In contrast to any other QD-based waveguide integrated SPS which has been demonstrated thus far~\cite{Kirsanske2017b,Liu2018}, our devices fulfill ultimate single photon purity and indistinguishability demands, imposed by boson sampling and linear optical quantum computing applications~\cite{Pan2012}. Performance of this source already outperforms any other on-chip integrated emitters including state-of-the-art heralded single photon spontaneous parametric down-conversion sources, where a maximum of 91\% photons indistinguishability has been achieved at 4-5\% source efficiency~\cite{Wang2016,Kaneda2018}. We believe that our device could be straightforwardly integrated with advanced on-chip functionalities including reconfigurable and reprogrammable optical circuits~\cite{Carolan2015} suitable for handling large scale multi-photon experiments. A potential of manufacturing such advanced quantum circuits combined with high purity indistinguishable SPS open a route towards fully integrated and thus scalable quantum information processing.

\begin{acknowledgments}
	The authors thank Silke Kuhn for fabricating the structures, and Dominick K\"{o}ck for calculating the electric field distribution in waveguides. \L{}.D. acknowledges the financial support from the Alexander von Humboldt Foundation. S.-H.K. acknowledges the financial support from the National Research Foundation of Korea through the Korean Government Grant NRF-2016R1C1B2007007. We are furthermore grateful for the support by the State of Bavaria.
\end{acknowledgments}

\bibliography{bib-manuscript}

\begin{thebibliography}{56}%
\makeatletter
\providecommand \@ifxundefined [1]{%
 \@ifx{#1\undefined}
}%
\providecommand \@ifnum [1]{%
 \ifnum #1\expandafter \@firstoftwo
 \else \expandafter \@secondoftwo
 \fi
}%
\providecommand \@ifx [1]{%
 \ifx #1\expandafter \@firstoftwo
 \else \expandafter \@secondoftwo
 \fi
}%
\providecommand \natexlab [1]{#1}%
\providecommand \enquote  [1]{``#1''}%
\providecommand \bibnamefont  [1]{#1}%
\providecommand \bibfnamefont [1]{#1}%
\providecommand \citenamefont [1]{#1}%
\providecommand \href@noop [0]{\@secondoftwo}%
\providecommand \href [0]{\begingroup \@sanitize@url \@href}%
\providecommand \@href[1]{\@@startlink{#1}\@@href}%
\providecommand \@@href[1]{\endgroup#1\@@endlink}%
\providecommand \@sanitize@url [0]{\catcode `\\12\catcode `\$12\catcode
  `\&12\catcode `\#12\catcode `\^12\catcode `\_12\catcode `\%12\relax}%
\providecommand \@@startlink[1]{}%
\providecommand \@@endlink[0]{}%
\providecommand \url  [0]{\begingroup\@sanitize@url \@url }%
\providecommand \@url [1]{\endgroup\@href {#1}{\urlprefix }}%
\providecommand \urlprefix  [0]{URL }%
\providecommand \Eprint [0]{\href }%
\providecommand \doibase [0]{http://dx.doi.org/}%
\providecommand \selectlanguage [0]{\@gobble}%
\providecommand \bibinfo  [0]{\@secondoftwo}%
\providecommand \bibfield  [0]{\@secondoftwo}%
\providecommand \translation [1]{[#1]}%
\providecommand \BibitemOpen [0]{}%
\providecommand \bibitemStop [0]{}%
\providecommand \bibitemNoStop [0]{.\EOS\space}%
\providecommand \EOS [0]{\spacefactor3000\relax}%
\providecommand \BibitemShut  [1]{\csname bibitem#1\endcsname}%
\let\auto@bib@innerbib\@empty
\bibitem [{\citenamefont {Kok}\ and\ \citenamefont
  {Lovett}(2010)}]{kok2010introduction}%
  \BibitemOpen
  \bibfield  {author} {\bibinfo {author} {\bibfnamefont {P.}~\bibnamefont
  {Kok}}\ and\ \bibinfo {author} {\bibfnamefont {B.}~\bibnamefont {Lovett}},\
  }\href {https://books.google.de/books?id=G2zKNooOeKcC} {\emph {\bibinfo
  {title} {Introduction to Optical Quantum Information Processing}}}\ (\bibinfo
   {publisher} {Cambridge University Press},\ \bibinfo {year}
  {2010})\BibitemShut {NoStop}%
\bibitem [{\citenamefont {Knill}\ \emph {et~al.}(2001)\citenamefont {Knill},
  \citenamefont {Laflamme},\ and\ \citenamefont {Milburn}}]{Knill2001}%
  \BibitemOpen
  \bibfield  {author} {\bibinfo {author} {\bibfnamefont {E.}~\bibnamefont
  {Knill}}, \bibinfo {author} {\bibfnamefont {R.}~\bibnamefont {Laflamme}}, \
  and\ \bibinfo {author} {\bibfnamefont {G.~J.}\ \bibnamefont {Milburn}},\
  }\href {\doibase 10.1038/35051009} {\bibfield  {journal} {\bibinfo  {journal}
  {Nature}\ }\textbf {\bibinfo {volume} {409}},\ \bibinfo {pages} {46}
  (\bibinfo {year} {2001})}\BibitemShut {NoStop}%
\bibitem [{\citenamefont {Spring}\ \emph {et~al.}(2013)\citenamefont {Spring},
  \citenamefont {Metcalf}, \citenamefont {Humphreys}, \citenamefont
  {Kolthammer}, \citenamefont {Jin}, \citenamefont {Barbieri}, \citenamefont
  {Datta}, \citenamefont {Thomas-Peter}, \citenamefont {Langford},
  \citenamefont {Kundys}, \citenamefont {Gates}, \citenamefont {Smith},
  \citenamefont {Smith},\ and\ \citenamefont {Walmsley}}]{Spring2013}%
  \BibitemOpen
  \bibfield  {author} {\bibinfo {author} {\bibfnamefont {J.~B.}\ \bibnamefont
  {Spring}}, \bibinfo {author} {\bibfnamefont {B.~J.}\ \bibnamefont {Metcalf}},
  \bibinfo {author} {\bibfnamefont {P.~C.}\ \bibnamefont {Humphreys}}, \bibinfo
  {author} {\bibfnamefont {W.~S.}\ \bibnamefont {Kolthammer}}, \bibinfo
  {author} {\bibfnamefont {X.-M.}\ \bibnamefont {Jin}}, \bibinfo {author}
  {\bibfnamefont {M.}~\bibnamefont {Barbieri}}, \bibinfo {author}
  {\bibfnamefont {A.}~\bibnamefont {Datta}}, \bibinfo {author} {\bibfnamefont
  {N.}~\bibnamefont {Thomas-Peter}}, \bibinfo {author} {\bibfnamefont {N.~K.}\
  \bibnamefont {Langford}}, \bibinfo {author} {\bibfnamefont {D.}~\bibnamefont
  {Kundys}}, \bibinfo {author} {\bibfnamefont {J.~C.}\ \bibnamefont {Gates}},
  \bibinfo {author} {\bibfnamefont {B.~J.}\ \bibnamefont {Smith}}, \bibinfo
  {author} {\bibfnamefont {P.~G.~R.}\ \bibnamefont {Smith}}, \ and\ \bibinfo
  {author} {\bibfnamefont {I.~A.}\ \bibnamefont {Walmsley}},\ }\href {\doibase
  10.1126/science.1231692} {\bibfield  {journal} {\bibinfo  {journal}
  {Science}\ }\textbf {\bibinfo {volume} {339}},\ \bibinfo {pages} {798}
  (\bibinfo {year} {2013})}\BibitemShut {NoStop}%
\bibitem [{\citenamefont {Briegel}\ \emph {et~al.}(1998)\citenamefont
  {Briegel}, \citenamefont {D{\"{u}}r}, \citenamefont {Cirac},\ and\
  \citenamefont {Zoller}}]{Briegel1998}%
  \BibitemOpen
  \bibfield  {author} {\bibinfo {author} {\bibfnamefont {H.-J.}\ \bibnamefont
  {Briegel}}, \bibinfo {author} {\bibfnamefont {W.}~\bibnamefont {D{\"{u}}r}},
  \bibinfo {author} {\bibfnamefont {J.~I.}\ \bibnamefont {Cirac}}, \ and\
  \bibinfo {author} {\bibfnamefont {P.}~\bibnamefont {Zoller}},\ }\href
  {\doibase 10.1103/PhysRevLett.81.5932} {\bibfield  {journal} {\bibinfo
  {journal} {Physical Review Letters}\ }\textbf {\bibinfo {volume} {81}},\
  \bibinfo {pages} {5932} (\bibinfo {year} {1998})}\BibitemShut {NoStop}%
\bibitem [{\citenamefont {Santori}\ \emph {et~al.}(2002)\citenamefont
  {Santori}, \citenamefont {Fattal}, \citenamefont {Vuckovi{\'{c}}},
  \citenamefont {Solomon},\ and\ \citenamefont {Yamamoto}}]{Santori2002}%
  \BibitemOpen
  \bibfield  {author} {\bibinfo {author} {\bibfnamefont {C.}~\bibnamefont
  {Santori}}, \bibinfo {author} {\bibfnamefont {D.}~\bibnamefont {Fattal}},
  \bibinfo {author} {\bibfnamefont {J.}~\bibnamefont {Vuckovi{\'{c}}}},
  \bibinfo {author} {\bibfnamefont {G.~S.}\ \bibnamefont {Solomon}}, \ and\
  \bibinfo {author} {\bibfnamefont {Y.}~\bibnamefont {Yamamoto}},\ }\href
  {\doibase 10.1038/419568a} {\bibfield  {journal} {\bibinfo  {journal}
  {Nature}\ }\textbf {\bibinfo {volume} {419}},\ \bibinfo {pages} {594}
  (\bibinfo {year} {2002})}\BibitemShut {NoStop}%
\bibitem [{\citenamefont {Aharonovich}\ \emph {et~al.}(2016)\citenamefont
  {Aharonovich}, \citenamefont {Englund},\ and\ \citenamefont
  {Toth}}]{Aharonovich2016}%
  \BibitemOpen
  \bibfield  {author} {\bibinfo {author} {\bibfnamefont {I.}~\bibnamefont
  {Aharonovich}}, \bibinfo {author} {\bibfnamefont {D.}~\bibnamefont
  {Englund}}, \ and\ \bibinfo {author} {\bibfnamefont {M.}~\bibnamefont
  {Toth}},\ }\href {\doibase 10.1038/nphoton.2016.186} {\bibfield  {journal}
  {\bibinfo  {journal} {Nature Photonics}\ }\textbf {\bibinfo {volume} {10}},\
  \bibinfo {pages} {631} (\bibinfo {year} {2016})}\BibitemShut {NoStop}%
\bibitem [{\citenamefont {Senellart}\ \emph {et~al.}(2017)\citenamefont
  {Senellart}, \citenamefont {Solomon},\ and\ \citenamefont
  {White}}]{Senellart2017}%
  \BibitemOpen
  \bibfield  {author} {\bibinfo {author} {\bibfnamefont {P.}~\bibnamefont
  {Senellart}}, \bibinfo {author} {\bibfnamefont {G.}~\bibnamefont {Solomon}},
  \ and\ \bibinfo {author} {\bibfnamefont {A.}~\bibnamefont {White}},\ }\href
  {\doibase 10.1038/nnano.2017.218} {\bibfield  {journal} {\bibinfo  {journal}
  {Nature Nanotechnology}\ }\textbf {\bibinfo {volume} {12}},\ \bibinfo {pages}
  {1026} (\bibinfo {year} {2017})}\BibitemShut {NoStop}%
\bibitem [{\citenamefont {Flagg}\ \emph {et~al.}(2009)\citenamefont {Flagg},
  \citenamefont {Muller}, \citenamefont {Robertson}, \citenamefont {Founta},
  \citenamefont {Deppe}, \citenamefont {Xiao}, \citenamefont {Ma},
  \citenamefont {Salamo},\ and\ \citenamefont {Shih}}]{Flagg2009}%
  \BibitemOpen
  \bibfield  {author} {\bibinfo {author} {\bibfnamefont {E.~B.}\ \bibnamefont
  {Flagg}}, \bibinfo {author} {\bibfnamefont {A.}~\bibnamefont {Muller}},
  \bibinfo {author} {\bibfnamefont {J.~W.}\ \bibnamefont {Robertson}}, \bibinfo
  {author} {\bibfnamefont {S.}~\bibnamefont {Founta}}, \bibinfo {author}
  {\bibfnamefont {D.~G.}\ \bibnamefont {Deppe}}, \bibinfo {author}
  {\bibfnamefont {M.}~\bibnamefont {Xiao}}, \bibinfo {author} {\bibfnamefont
  {W.}~\bibnamefont {Ma}}, \bibinfo {author} {\bibfnamefont {G.~J.}\
  \bibnamefont {Salamo}}, \ and\ \bibinfo {author} {\bibfnamefont {C.~K.}\
  \bibnamefont {Shih}},\ }\href {\doibase 10.1038/nphys1184} {\bibfield
  {journal} {\bibinfo  {journal} {Nature Physics}\ }\textbf {\bibinfo {volume}
  {5}},\ \bibinfo {pages} {203} (\bibinfo {year} {2009})}\BibitemShut {NoStop}%
\bibitem [{\citenamefont {Ates}\ \emph {et~al.}(2009)\citenamefont {Ates},
  \citenamefont {Ulrich}, \citenamefont {Reitzenstein}, \citenamefont
  {L{\"{o}}ffler}, \citenamefont {Forchel},\ and\ \citenamefont
  {Michler}}]{Ates2009}%
  \BibitemOpen
  \bibfield  {author} {\bibinfo {author} {\bibfnamefont {S.}~\bibnamefont
  {Ates}}, \bibinfo {author} {\bibfnamefont {S.~M.}\ \bibnamefont {Ulrich}},
  \bibinfo {author} {\bibfnamefont {S.}~\bibnamefont {Reitzenstein}}, \bibinfo
  {author} {\bibfnamefont {A.}~\bibnamefont {L{\"{o}}ffler}}, \bibinfo {author}
  {\bibfnamefont {A.}~\bibnamefont {Forchel}}, \ and\ \bibinfo {author}
  {\bibfnamefont {P.}~\bibnamefont {Michler}},\ }\href {\doibase
  10.1103/PhysRevLett.103.167402} {\bibfield  {journal} {\bibinfo  {journal}
  {Physical Review Letters}\ }\textbf {\bibinfo {volume} {103}},\ \bibinfo
  {pages} {167402} (\bibinfo {year} {2009})}\BibitemShut {NoStop}%
\bibitem [{\citenamefont {Ding}\ \emph {et~al.}(2016)\citenamefont {Ding},
  \citenamefont {He}, \citenamefont {Duan}, \citenamefont {Gregersen},
  \citenamefont {Chen}, \citenamefont {Unsleber}, \citenamefont {Maier},
  \citenamefont {Schneider}, \citenamefont {Kamp}, \citenamefont
  {H{\"{o}}fling}, \citenamefont {Lu},\ and\ \citenamefont {Pan}}]{Ding2016}%
  \BibitemOpen
  \bibfield  {author} {\bibinfo {author} {\bibfnamefont {X.}~\bibnamefont
  {Ding}}, \bibinfo {author} {\bibfnamefont {Y.}~\bibnamefont {He}}, \bibinfo
  {author} {\bibfnamefont {Z.~C.}\ \bibnamefont {Duan}}, \bibinfo {author}
  {\bibfnamefont {N.}~\bibnamefont {Gregersen}}, \bibinfo {author}
  {\bibfnamefont {M.~C.}\ \bibnamefont {Chen}}, \bibinfo {author}
  {\bibfnamefont {S.}~\bibnamefont {Unsleber}}, \bibinfo {author}
  {\bibfnamefont {S.}~\bibnamefont {Maier}}, \bibinfo {author} {\bibfnamefont
  {C.}~\bibnamefont {Schneider}}, \bibinfo {author} {\bibfnamefont
  {M.}~\bibnamefont {Kamp}}, \bibinfo {author} {\bibfnamefont {S.}~\bibnamefont
  {H{\"{o}}fling}}, \bibinfo {author} {\bibfnamefont {C.-Y.}\ \bibnamefont
  {Lu}}, \ and\ \bibinfo {author} {\bibfnamefont {J.-W.}\ \bibnamefont {Pan}},\
  }\href {\doibase 10.1103/PhysRevLett.116.020401} {\bibfield  {journal}
  {\bibinfo  {journal} {Physical Review Letters}\ }\textbf {\bibinfo {volume}
  {116}},\ \bibinfo {pages} {020401} (\bibinfo {year} {2016})}\BibitemShut
  {NoStop}%
\bibitem [{\citenamefont {Somaschi}\ \emph {et~al.}(2016)\citenamefont
  {Somaschi}, \citenamefont {Giesz}, \citenamefont {{De Santis}}, \citenamefont
  {Loredo}, \citenamefont {Almeida}, \citenamefont {Hornecker}, \citenamefont
  {Portalupi}, \citenamefont {Grange}, \citenamefont {Ant{\'{o}}n},
  \citenamefont {Demory}, \citenamefont {G{\'{o}}mez}, \citenamefont {Sagnes},
  \citenamefont {Lanzillotti-Kimura}, \citenamefont {Lema{\'{i}}tre},
  \citenamefont {Auffeves}, \citenamefont {White}, \citenamefont {Lanco},\ and\
  \citenamefont {Senellart}}]{Somaschi2016}%
  \BibitemOpen
  \bibfield  {author} {\bibinfo {author} {\bibfnamefont {N.}~\bibnamefont
  {Somaschi}}, \bibinfo {author} {\bibfnamefont {V.}~\bibnamefont {Giesz}},
  \bibinfo {author} {\bibfnamefont {L.}~\bibnamefont {{De Santis}}}, \bibinfo
  {author} {\bibfnamefont {J.~C.}\ \bibnamefont {Loredo}}, \bibinfo {author}
  {\bibfnamefont {M.~P.}\ \bibnamefont {Almeida}}, \bibinfo {author}
  {\bibfnamefont {G.}~\bibnamefont {Hornecker}}, \bibinfo {author}
  {\bibfnamefont {S.~L.}\ \bibnamefont {Portalupi}}, \bibinfo {author}
  {\bibfnamefont {T.}~\bibnamefont {Grange}}, \bibinfo {author} {\bibfnamefont
  {C.}~\bibnamefont {Ant{\'{o}}n}}, \bibinfo {author} {\bibfnamefont
  {J.}~\bibnamefont {Demory}}, \bibinfo {author} {\bibfnamefont
  {C.}~\bibnamefont {G{\'{o}}mez}}, \bibinfo {author} {\bibfnamefont
  {I.}~\bibnamefont {Sagnes}}, \bibinfo {author} {\bibfnamefont {N.~D.}\
  \bibnamefont {Lanzillotti-Kimura}}, \bibinfo {author} {\bibfnamefont
  {A.}~\bibnamefont {Lema{\'{i}}tre}}, \bibinfo {author} {\bibfnamefont
  {A.}~\bibnamefont {Auffeves}}, \bibinfo {author} {\bibfnamefont {A.~G.}\
  \bibnamefont {White}}, \bibinfo {author} {\bibfnamefont {L.}~\bibnamefont
  {Lanco}}, \ and\ \bibinfo {author} {\bibfnamefont {P.}~\bibnamefont
  {Senellart}},\ }\href {\doibase 10.1038/nphoton.2016.23} {\bibfield
  {journal} {\bibinfo  {journal} {Nature Photonics}\ }\textbf {\bibinfo
  {volume} {10}},\ \bibinfo {pages} {340} (\bibinfo {year} {2016})}\BibitemShut
  {NoStop}%
\bibitem [{\citenamefont {Unsleber}\ \emph {et~al.}(2016)\citenamefont
  {Unsleber}, \citenamefont {He}, \citenamefont {Maier}, \citenamefont
  {Gerhardt}, \citenamefont {Lu}, \citenamefont {Pan}, \citenamefont {Kamp},
  \citenamefont {Schneider},\ and\ \citenamefont
  {H{\"{o}}fling}}]{Unsleber2016}%
  \BibitemOpen
  \bibfield  {author} {\bibinfo {author} {\bibfnamefont {S.}~\bibnamefont
  {Unsleber}}, \bibinfo {author} {\bibfnamefont {Y.-M.}\ \bibnamefont {He}},
  \bibinfo {author} {\bibfnamefont {S.}~\bibnamefont {Maier}}, \bibinfo
  {author} {\bibfnamefont {S.}~\bibnamefont {Gerhardt}}, \bibinfo {author}
  {\bibfnamefont {C.-Y.}\ \bibnamefont {Lu}}, \bibinfo {author} {\bibfnamefont
  {J.-W.}\ \bibnamefont {Pan}}, \bibinfo {author} {\bibfnamefont
  {M.}~\bibnamefont {Kamp}}, \bibinfo {author} {\bibfnamefont {C.}~\bibnamefont
  {Schneider}}, \ and\ \bibinfo {author} {\bibfnamefont {S.}~\bibnamefont
  {H{\"{o}}fling}},\ }\href {\doibase 10.1364/OE.24.008539} {\bibfield
  {journal} {\bibinfo  {journal} {Optics express}\ }\textbf {\bibinfo {volume}
  {24}},\ \bibinfo {pages} {8539} (\bibinfo {year} {2016})}\BibitemShut
  {NoStop}%
\bibitem [{\citenamefont {Fischer}\ \emph {et~al.}(2016)\citenamefont
  {Fischer}, \citenamefont {M{\"{u}}ller}, \citenamefont {Lagoudakis},\ and\
  \citenamefont {Vu{\v{c}}kovi{\'{c}}}}]{Fischer2016}%
  \BibitemOpen
  \bibfield  {author} {\bibinfo {author} {\bibfnamefont {K.~A.}\ \bibnamefont
  {Fischer}}, \bibinfo {author} {\bibfnamefont {K.}~\bibnamefont
  {M{\"{u}}ller}}, \bibinfo {author} {\bibfnamefont {K.~G.}\ \bibnamefont
  {Lagoudakis}}, \ and\ \bibinfo {author} {\bibfnamefont {J.}~\bibnamefont
  {Vu{\v{c}}kovi{\'{c}}}},\ }\href {\doibase 10.1088/1367-2630/18/11/113053}
  {\bibfield  {journal} {\bibinfo  {journal} {New Journal of Physics}\ }\textbf
  {\bibinfo {volume} {18}},\ \bibinfo {pages} {113053} (\bibinfo {year}
  {2016})}\BibitemShut {NoStop}%
\bibitem [{\citenamefont {Zhang}\ \emph {et~al.}(2016)\citenamefont {Zhang},
  \citenamefont {Chen}, \citenamefont {Mietschke}, \citenamefont {Zhang},
  \citenamefont {Yuan}, \citenamefont {Abel}, \citenamefont {H{\"{u}}hne},
  \citenamefont {Nielsch}, \citenamefont {Fompeyrine}, \citenamefont {Ding},\
  and\ \citenamefont {Schmidt}}]{Zhang2016}%
  \BibitemOpen
  \bibfield  {author} {\bibinfo {author} {\bibfnamefont {Y.}~\bibnamefont
  {Zhang}}, \bibinfo {author} {\bibfnamefont {Y.}~\bibnamefont {Chen}},
  \bibinfo {author} {\bibfnamefont {M.}~\bibnamefont {Mietschke}}, \bibinfo
  {author} {\bibfnamefont {L.}~\bibnamefont {Zhang}}, \bibinfo {author}
  {\bibfnamefont {F.}~\bibnamefont {Yuan}}, \bibinfo {author} {\bibfnamefont
  {S.}~\bibnamefont {Abel}}, \bibinfo {author} {\bibfnamefont {R.}~\bibnamefont
  {H{\"{u}}hne}}, \bibinfo {author} {\bibfnamefont {K.}~\bibnamefont
  {Nielsch}}, \bibinfo {author} {\bibfnamefont {J.}~\bibnamefont {Fompeyrine}},
  \bibinfo {author} {\bibfnamefont {F.}~\bibnamefont {Ding}}, \ and\ \bibinfo
  {author} {\bibfnamefont {O.~G.}\ \bibnamefont {Schmidt}},\ }\href {\doibase
  10.1021/acs.nanolett.6b02523} {\bibfield  {journal} {\bibinfo  {journal}
  {Nano Letters}\ }\textbf {\bibinfo {volume} {16}},\ \bibinfo {pages} {5785}
  (\bibinfo {year} {2016})}\BibitemShut {NoStop}%
\bibitem [{\citenamefont {Trotta}\ \emph {et~al.}(2016)\citenamefont {Trotta},
  \citenamefont {Mart{\'{i}}n-S{\'{a}}nchez}, \citenamefont {Wildmann},
  \citenamefont {Piredda}, \citenamefont {Reindl}, \citenamefont {Schimpf},
  \citenamefont {Zallo}, \citenamefont {Stroj}, \citenamefont {Edlinger},\ and\
  \citenamefont {Rastelli}}]{Trotta2016}%
  \BibitemOpen
  \bibfield  {author} {\bibinfo {author} {\bibfnamefont {R.}~\bibnamefont
  {Trotta}}, \bibinfo {author} {\bibfnamefont {J.}~\bibnamefont
  {Mart{\'{i}}n-S{\'{a}}nchez}}, \bibinfo {author} {\bibfnamefont {J.~S.}\
  \bibnamefont {Wildmann}}, \bibinfo {author} {\bibfnamefont {G.}~\bibnamefont
  {Piredda}}, \bibinfo {author} {\bibfnamefont {M.}~\bibnamefont {Reindl}},
  \bibinfo {author} {\bibfnamefont {C.}~\bibnamefont {Schimpf}}, \bibinfo
  {author} {\bibfnamefont {E.}~\bibnamefont {Zallo}}, \bibinfo {author}
  {\bibfnamefont {S.}~\bibnamefont {Stroj}}, \bibinfo {author} {\bibfnamefont
  {J.}~\bibnamefont {Edlinger}}, \ and\ \bibinfo {author} {\bibfnamefont
  {A.}~\bibnamefont {Rastelli}},\ }\href {\doibase 10.1038/ncomms10375}
  {\bibfield  {journal} {\bibinfo  {journal} {Nature Communications}\ }\textbf
  {\bibinfo {volume} {7}},\ \bibinfo {pages} {10375} (\bibinfo {year}
  {2016})}\BibitemShut {NoStop}%
\bibitem [{\citenamefont {Kir{\v{s}}ansk{\.e}}\ \emph
  {et~al.}(2017)\citenamefont {Kir{\v{s}}ansk{\.e}}, \citenamefont
  {Thyrrestrup}, \citenamefont {Daveau}, \citenamefont {Dree{\ss}en},
  \citenamefont {Pregnolato}, \citenamefont {Midolo}, \citenamefont
  {Tighineanu}, \citenamefont {Javadi}, \citenamefont {Stobbe}, \citenamefont
  {Schott}, \citenamefont {Ludwig}, \citenamefont {Wieck}, \citenamefont
  {Park}, \citenamefont {Song}, \citenamefont {Kuhlmann}, \citenamefont
  {S{\"{o}}llner}, \citenamefont {L{\"{o}}bl}, \citenamefont {Warburton},\ and\
  \citenamefont {Lodahl}}]{Kirsanske2017b}%
  \BibitemOpen
  \bibfield  {author} {\bibinfo {author} {\bibfnamefont {G.}~\bibnamefont
  {Kir{\v{s}}ansk{\.e}}}, \bibinfo {author} {\bibfnamefont {H.}~\bibnamefont
  {Thyrrestrup}}, \bibinfo {author} {\bibfnamefont {R.~S.}\ \bibnamefont
  {Daveau}}, \bibinfo {author} {\bibfnamefont {C.~L.}\ \bibnamefont
  {Dree{\ss}en}}, \bibinfo {author} {\bibfnamefont {T.}~\bibnamefont
  {Pregnolato}}, \bibinfo {author} {\bibfnamefont {L.}~\bibnamefont {Midolo}},
  \bibinfo {author} {\bibfnamefont {P.}~\bibnamefont {Tighineanu}}, \bibinfo
  {author} {\bibfnamefont {A.}~\bibnamefont {Javadi}}, \bibinfo {author}
  {\bibfnamefont {S.}~\bibnamefont {Stobbe}}, \bibinfo {author} {\bibfnamefont
  {R.}~\bibnamefont {Schott}}, \bibinfo {author} {\bibfnamefont
  {A.}~\bibnamefont {Ludwig}}, \bibinfo {author} {\bibfnamefont {A.~D.}\
  \bibnamefont {Wieck}}, \bibinfo {author} {\bibfnamefont {S.~I.}\ \bibnamefont
  {Park}}, \bibinfo {author} {\bibfnamefont {J.~D.}\ \bibnamefont {Song}},
  \bibinfo {author} {\bibfnamefont {A.~V.}\ \bibnamefont {Kuhlmann}}, \bibinfo
  {author} {\bibfnamefont {I.}~\bibnamefont {S{\"{o}}llner}}, \bibinfo {author}
  {\bibfnamefont {M.~C.}\ \bibnamefont {L{\"{o}}bl}}, \bibinfo {author}
  {\bibfnamefont {R.~J.}\ \bibnamefont {Warburton}}, \ and\ \bibinfo {author}
  {\bibfnamefont {P.}~\bibnamefont {Lodahl}},\ }\href {\doibase
  10.1103/PhysRevB.96.165306} {\bibfield  {journal} {\bibinfo  {journal}
  {Physical Review B}\ }\textbf {\bibinfo {volume} {96}},\ \bibinfo {pages}
  {165306} (\bibinfo {year} {2017})}\BibitemShut {NoStop}%
\bibitem [{\citenamefont {Hoang}\ \emph {et~al.}(2012)\citenamefont {Hoang},
  \citenamefont {Beetz}, \citenamefont {Lermer}, \citenamefont {Midolo},
  \citenamefont {Kamp}, \citenamefont {H{\"{o}}fling},\ and\ \citenamefont
  {Fiore}}]{Hoang2012}%
  \BibitemOpen
  \bibfield  {author} {\bibinfo {author} {\bibfnamefont {T.~B.}\ \bibnamefont
  {Hoang}}, \bibinfo {author} {\bibfnamefont {J.}~\bibnamefont {Beetz}},
  \bibinfo {author} {\bibfnamefont {M.}~\bibnamefont {Lermer}}, \bibinfo
  {author} {\bibfnamefont {L.}~\bibnamefont {Midolo}}, \bibinfo {author}
  {\bibfnamefont {M.}~\bibnamefont {Kamp}}, \bibinfo {author} {\bibfnamefont
  {S.}~\bibnamefont {H{\"{o}}fling}}, \ and\ \bibinfo {author} {\bibfnamefont
  {A.}~\bibnamefont {Fiore}},\ }\href {\doibase 10.1364/OE.20.021758}
  {\bibfield  {journal} {\bibinfo  {journal} {Optics Express}\ }\textbf
  {\bibinfo {volume} {20}},\ \bibinfo {pages} {21758} (\bibinfo {year}
  {2012})}\BibitemShut {NoStop}%
\bibitem [{\citenamefont {Pooley}\ \emph {et~al.}(2012)\citenamefont {Pooley},
  \citenamefont {Ellis}, \citenamefont {Patel}, \citenamefont {Bennett},
  \citenamefont {Chan}, \citenamefont {Farrer}, \citenamefont {Ritchie},\ and\
  \citenamefont {Shields}}]{Pooley2012}%
  \BibitemOpen
  \bibfield  {author} {\bibinfo {author} {\bibfnamefont {M.~A.}\ \bibnamefont
  {Pooley}}, \bibinfo {author} {\bibfnamefont {D.~J.~P.}\ \bibnamefont
  {Ellis}}, \bibinfo {author} {\bibfnamefont {R.~B.}\ \bibnamefont {Patel}},
  \bibinfo {author} {\bibfnamefont {A.~J.}\ \bibnamefont {Bennett}}, \bibinfo
  {author} {\bibfnamefont {K.~H.~A.}\ \bibnamefont {Chan}}, \bibinfo {author}
  {\bibfnamefont {I.}~\bibnamefont {Farrer}}, \bibinfo {author} {\bibfnamefont
  {D.~A.}\ \bibnamefont {Ritchie}}, \ and\ \bibinfo {author} {\bibfnamefont
  {A.~J.}\ \bibnamefont {Shields}},\ }\href {\doibase 10.1063/1.4719077}
  {\bibfield  {journal} {\bibinfo  {journal} {Applied Physics Letters}\
  }\textbf {\bibinfo {volume} {100}},\ \bibinfo {pages} {211103} (\bibinfo
  {year} {2012})}\BibitemShut {NoStop}%
\bibitem [{\citenamefont {Gazzano}\ \emph {et~al.}(2013)\citenamefont
  {Gazzano}, \citenamefont {Almeida}, \citenamefont {Nowak}, \citenamefont
  {Portalupi}, \citenamefont {Lema{\^{i}}tre}, \citenamefont {Sagnes},
  \citenamefont {White},\ and\ \citenamefont {Senellart}}]{Gazzano2013a}%
  \BibitemOpen
  \bibfield  {author} {\bibinfo {author} {\bibfnamefont {O.}~\bibnamefont
  {Gazzano}}, \bibinfo {author} {\bibfnamefont {M.~P.}\ \bibnamefont
  {Almeida}}, \bibinfo {author} {\bibfnamefont {A.~K.}\ \bibnamefont {Nowak}},
  \bibinfo {author} {\bibfnamefont {S.~L.}\ \bibnamefont {Portalupi}}, \bibinfo
  {author} {\bibfnamefont {A.}~\bibnamefont {Lema{\^{i}}tre}}, \bibinfo
  {author} {\bibfnamefont {I.}~\bibnamefont {Sagnes}}, \bibinfo {author}
  {\bibfnamefont {A.~G.}\ \bibnamefont {White}}, \ and\ \bibinfo {author}
  {\bibfnamefont {P.}~\bibnamefont {Senellart}},\ }\href {\doibase
  10.1103/PhysRevLett.110.250501} {\bibfield  {journal} {\bibinfo  {journal}
  {Physical Review Letters}\ }\textbf {\bibinfo {volume} {110}},\ \bibinfo
  {pages} {250501} (\bibinfo {year} {2013})}\BibitemShut {NoStop}%
\bibitem [{\citenamefont {He}\ \emph {et~al.}(2013)\citenamefont {He},
  \citenamefont {He}, \citenamefont {Wei}, \citenamefont {Wu}, \citenamefont
  {Atat{\"{u}}re}, \citenamefont {Schneider}, \citenamefont {H{\"{o}}fling},
  \citenamefont {Kamp}, \citenamefont {Lu},\ and\ \citenamefont
  {Pan}}]{He2013a}%
  \BibitemOpen
  \bibfield  {author} {\bibinfo {author} {\bibfnamefont {Y.-M.}\ \bibnamefont
  {He}}, \bibinfo {author} {\bibfnamefont {Y.}~\bibnamefont {He}}, \bibinfo
  {author} {\bibfnamefont {Y.-J.}\ \bibnamefont {Wei}}, \bibinfo {author}
  {\bibfnamefont {D.}~\bibnamefont {Wu}}, \bibinfo {author} {\bibfnamefont
  {M.}~\bibnamefont {Atat{\"{u}}re}}, \bibinfo {author} {\bibfnamefont
  {C.}~\bibnamefont {Schneider}}, \bibinfo {author} {\bibfnamefont
  {S.}~\bibnamefont {H{\"{o}}fling}}, \bibinfo {author} {\bibfnamefont
  {M.}~\bibnamefont {Kamp}}, \bibinfo {author} {\bibfnamefont {C.-Y.}\
  \bibnamefont {Lu}}, \ and\ \bibinfo {author} {\bibfnamefont {J.-W.}\
  \bibnamefont {Pan}},\ }\href {\doibase 10.1038/nnano.2012.262} {\bibfield
  {journal} {\bibinfo  {journal} {Nature Nanotechnology}\ }\textbf {\bibinfo
  {volume} {8}},\ \bibinfo {pages} {213} (\bibinfo {year} {2013})}\BibitemShut
  {NoStop}%
\bibitem [{\citenamefont {Delteil}\ \emph {et~al.}(2015)\citenamefont
  {Delteil}, \citenamefont {Sun}, \citenamefont {Gao}, \citenamefont {Togan},
  \citenamefont {Faelt},\ and\ \citenamefont {Imamoğlu}}]{Delteil2015}%
  \BibitemOpen
  \bibfield  {author} {\bibinfo {author} {\bibfnamefont {A.}~\bibnamefont
  {Delteil}}, \bibinfo {author} {\bibfnamefont {Z.}~\bibnamefont {Sun}},
  \bibinfo {author} {\bibfnamefont {W.-b.}\ \bibnamefont {Gao}}, \bibinfo
  {author} {\bibfnamefont {E.}~\bibnamefont {Togan}}, \bibinfo {author}
  {\bibfnamefont {S.}~\bibnamefont {Faelt}}, \ and\ \bibinfo {author}
  {\bibfnamefont {A.}~\bibnamefont {Imamoğlu}},\ }\href {\doibase
  10.1038/nphys3605} {\bibfield  {journal} {\bibinfo  {journal} {Nature
  Physics}\ }\textbf {\bibinfo {volume} {12}},\ \bibinfo {pages} {218}
  (\bibinfo {year} {2015})}\BibitemShut {NoStop}%
\bibitem [{\citenamefont {Wang}\ \emph {et~al.}(2017)\citenamefont {Wang},
  \citenamefont {He}, \citenamefont {Li}, \citenamefont {Su}, \citenamefont
  {Li}, \citenamefont {Huang}, \citenamefont {Ding}, \citenamefont {Chen},
  \citenamefont {Liu}, \citenamefont {Qin}, \citenamefont {Li}, \citenamefont
  {He}, \citenamefont {Schneider}, \citenamefont {Kamp}, \citenamefont {Peng},
  \citenamefont {H{\"{o}}fling}, \citenamefont {Lu},\ and\ \citenamefont
  {Pan}}]{Wang2017}%
  \BibitemOpen
  \bibfield  {author} {\bibinfo {author} {\bibfnamefont {H.}~\bibnamefont
  {Wang}}, \bibinfo {author} {\bibfnamefont {Y.-M.}\ \bibnamefont {He}},
  \bibinfo {author} {\bibfnamefont {Y.-H.}\ \bibnamefont {Li}}, \bibinfo
  {author} {\bibfnamefont {Z.-E.}\ \bibnamefont {Su}}, \bibinfo {author}
  {\bibfnamefont {B.}~\bibnamefont {Li}}, \bibinfo {author} {\bibfnamefont
  {H.-L.}\ \bibnamefont {Huang}}, \bibinfo {author} {\bibfnamefont
  {X.}~\bibnamefont {Ding}}, \bibinfo {author} {\bibfnamefont {M.-C.}\
  \bibnamefont {Chen}}, \bibinfo {author} {\bibfnamefont {C.}~\bibnamefont
  {Liu}}, \bibinfo {author} {\bibfnamefont {J.}~\bibnamefont {Qin}}, \bibinfo
  {author} {\bibfnamefont {J.-P.}\ \bibnamefont {Li}}, \bibinfo {author}
  {\bibfnamefont {Y.-M.}\ \bibnamefont {He}}, \bibinfo {author} {\bibfnamefont
  {C.}~\bibnamefont {Schneider}}, \bibinfo {author} {\bibfnamefont
  {M.}~\bibnamefont {Kamp}}, \bibinfo {author} {\bibfnamefont {C.-Z.}\
  \bibnamefont {Peng}}, \bibinfo {author} {\bibfnamefont {S.}~\bibnamefont
  {H{\"{o}}fling}}, \bibinfo {author} {\bibfnamefont {C.-Y.}\ \bibnamefont
  {Lu}}, \ and\ \bibinfo {author} {\bibfnamefont {J.-W.}\ \bibnamefont {Pan}},\
  }\href {\doibase 10.1038/nphoton.2017.63} {\bibfield  {journal} {\bibinfo
  {journal} {Nature Photonics}\ }\textbf {\bibinfo {volume} {11}},\ \bibinfo
  {pages} {361} (\bibinfo {year} {2017})}\BibitemShut {NoStop}%
\bibitem [{\citenamefont {Loredo}\ \emph {et~al.}(2017)\citenamefont {Loredo},
  \citenamefont {Broome}, \citenamefont {Hilaire}, \citenamefont {Gazzano},
  \citenamefont {Sagnes}, \citenamefont {Lemaitre}, \citenamefont {Almeida},
  \citenamefont {Senellart},\ and\ \citenamefont {White}}]{Loredo2017}%
  \BibitemOpen
  \bibfield  {author} {\bibinfo {author} {\bibfnamefont {J.~C.}\ \bibnamefont
  {Loredo}}, \bibinfo {author} {\bibfnamefont {M.~A.}\ \bibnamefont {Broome}},
  \bibinfo {author} {\bibfnamefont {P.}~\bibnamefont {Hilaire}}, \bibinfo
  {author} {\bibfnamefont {O.}~\bibnamefont {Gazzano}}, \bibinfo {author}
  {\bibfnamefont {I.}~\bibnamefont {Sagnes}}, \bibinfo {author} {\bibfnamefont
  {A.}~\bibnamefont {Lemaitre}}, \bibinfo {author} {\bibfnamefont {M.~P.}\
  \bibnamefont {Almeida}}, \bibinfo {author} {\bibfnamefont {P.}~\bibnamefont
  {Senellart}}, \ and\ \bibinfo {author} {\bibfnamefont {A.~G.}\ \bibnamefont
  {White}},\ }\href {\doibase 10.1103/PhysRevLett.118.130503} {\bibfield
  {journal} {\bibinfo  {journal} {Physical Review Letters}\ }\textbf {\bibinfo
  {volume} {118}},\ \bibinfo {pages} {130503} (\bibinfo {year}
  {2017})}\BibitemShut {NoStop}%
\bibitem [{\citenamefont {Dietrich}\ \emph {et~al.}(2016)\citenamefont
  {Dietrich}, \citenamefont {Fiore}, \citenamefont {Thompson}, \citenamefont
  {Kamp},\ and\ \citenamefont {H{\"{o}}fling}}]{Dietrich2016a}%
  \BibitemOpen
  \bibfield  {author} {\bibinfo {author} {\bibfnamefont {C.~P.}\ \bibnamefont
  {Dietrich}}, \bibinfo {author} {\bibfnamefont {A.}~\bibnamefont {Fiore}},
  \bibinfo {author} {\bibfnamefont {M.~G.}\ \bibnamefont {Thompson}}, \bibinfo
  {author} {\bibfnamefont {M.}~\bibnamefont {Kamp}}, \ and\ \bibinfo {author}
  {\bibfnamefont {S.}~\bibnamefont {H{\"{o}}fling}},\ }\href {\doibase
  10.1002/lpor.201500321} {\bibfield  {journal} {\bibinfo  {journal} {Laser
  {\&} Photonics Reviews}\ }\textbf {\bibinfo {volume} {10}},\ \bibinfo {pages}
  {870} (\bibinfo {year} {2016})}\BibitemShut {NoStop}%
\bibitem [{\citenamefont {J{\"{o}}ns}\ \emph {et~al.}(2015)\citenamefont
  {J{\"{o}}ns}, \citenamefont {Rengstl}, \citenamefont {Oster}, \citenamefont
  {Hargart}, \citenamefont {Heldmaier}, \citenamefont {Bounouar}, \citenamefont
  {Ulrich}, \citenamefont {Jetter},\ and\ \citenamefont {Michler}}]{Jons2015}%
  \BibitemOpen
  \bibfield  {author} {\bibinfo {author} {\bibfnamefont {K.~D.}\ \bibnamefont
  {J{\"{o}}ns}}, \bibinfo {author} {\bibfnamefont {U.}~\bibnamefont {Rengstl}},
  \bibinfo {author} {\bibfnamefont {M.}~\bibnamefont {Oster}}, \bibinfo
  {author} {\bibfnamefont {F.}~\bibnamefont {Hargart}}, \bibinfo {author}
  {\bibfnamefont {M.}~\bibnamefont {Heldmaier}}, \bibinfo {author}
  {\bibfnamefont {S.}~\bibnamefont {Bounouar}}, \bibinfo {author}
  {\bibfnamefont {S.~M.}\ \bibnamefont {Ulrich}}, \bibinfo {author}
  {\bibfnamefont {M.}~\bibnamefont {Jetter}}, \ and\ \bibinfo {author}
  {\bibfnamefont {P.}~\bibnamefont {Michler}},\ }\href {\doibase
  10.1088/0022-3727/48/8/085101} {\bibfield  {journal} {\bibinfo  {journal}
  {Journal of Physics D: Applied Physics}\ }\textbf {\bibinfo {volume} {48}},\
  \bibinfo {pages} {085101} (\bibinfo {year} {2015})}\BibitemShut {NoStop}%
\bibitem [{\citenamefont {Enderlin}\ \emph {et~al.}(2012)\citenamefont
  {Enderlin}, \citenamefont {Ota}, \citenamefont {Ohta}, \citenamefont
  {Kumagai}, \citenamefont {Ishida}, \citenamefont {Iwamoto},\ and\
  \citenamefont {Arakawa}}]{Enderlin2012}%
  \BibitemOpen
  \bibfield  {author} {\bibinfo {author} {\bibfnamefont {A.}~\bibnamefont
  {Enderlin}}, \bibinfo {author} {\bibfnamefont {Y.}~\bibnamefont {Ota}},
  \bibinfo {author} {\bibfnamefont {R.}~\bibnamefont {Ohta}}, \bibinfo {author}
  {\bibfnamefont {N.}~\bibnamefont {Kumagai}}, \bibinfo {author} {\bibfnamefont
  {S.}~\bibnamefont {Ishida}}, \bibinfo {author} {\bibfnamefont
  {S.}~\bibnamefont {Iwamoto}}, \ and\ \bibinfo {author} {\bibfnamefont
  {Y.}~\bibnamefont {Arakawa}},\ }\href {\doibase 10.1103/PhysRevB.86.075314}
  {\bibfield  {journal} {\bibinfo  {journal} {Physical Review B}\ }\textbf
  {\bibinfo {volume} {86}},\ \bibinfo {pages} {075314} (\bibinfo {year}
  {2012})}\BibitemShut {NoStop}%
\bibitem [{\citenamefont {Schwagmann}\ \emph {et~al.}(2011)\citenamefont
  {Schwagmann}, \citenamefont {Kalliakos}, \citenamefont {Farrer},
  \citenamefont {Griffiths}, \citenamefont {Jones}, \citenamefont {Ritchie},\
  and\ \citenamefont {Shields}}]{Schwagmann2011}%
  \BibitemOpen
  \bibfield  {author} {\bibinfo {author} {\bibfnamefont {A.}~\bibnamefont
  {Schwagmann}}, \bibinfo {author} {\bibfnamefont {S.}~\bibnamefont
  {Kalliakos}}, \bibinfo {author} {\bibfnamefont {I.}~\bibnamefont {Farrer}},
  \bibinfo {author} {\bibfnamefont {J.~P.}\ \bibnamefont {Griffiths}}, \bibinfo
  {author} {\bibfnamefont {G.~A.~C.}\ \bibnamefont {Jones}}, \bibinfo {author}
  {\bibfnamefont {D.~A.}\ \bibnamefont {Ritchie}}, \ and\ \bibinfo {author}
  {\bibfnamefont {A.~J.}\ \bibnamefont {Shields}},\ }\href {\doibase
  10.1063/1.3672214} {\bibfield  {journal} {\bibinfo  {journal} {Applied
  Physics Letters}\ }\textbf {\bibinfo {volume} {99}},\ \bibinfo {pages}
  {261108} (\bibinfo {year} {2011})}\BibitemShut {NoStop}%
\bibitem [{\citenamefont {Arcari}\ \emph {et~al.}(2014)\citenamefont {Arcari},
  \citenamefont {S{\"{o}}llner}, \citenamefont {Javadi}, \citenamefont
  {{Lindskov Hansen}}, \citenamefont {Mahmoodian}, \citenamefont {Liu},
  \citenamefont {Thyrrestrup}, \citenamefont {Lee}, \citenamefont {Song},
  \citenamefont {Stobbe},\ and\ \citenamefont {Lodahl}}]{Arcari2014}%
  \BibitemOpen
  \bibfield  {author} {\bibinfo {author} {\bibfnamefont {M.}~\bibnamefont
  {Arcari}}, \bibinfo {author} {\bibfnamefont {I.}~\bibnamefont
  {S{\"{o}}llner}}, \bibinfo {author} {\bibfnamefont {A.}~\bibnamefont
  {Javadi}}, \bibinfo {author} {\bibfnamefont {S.}~\bibnamefont {{Lindskov
  Hansen}}}, \bibinfo {author} {\bibfnamefont {S.}~\bibnamefont {Mahmoodian}},
  \bibinfo {author} {\bibfnamefont {J.}~\bibnamefont {Liu}}, \bibinfo {author}
  {\bibfnamefont {H.}~\bibnamefont {Thyrrestrup}}, \bibinfo {author}
  {\bibfnamefont {E.~H.}\ \bibnamefont {Lee}}, \bibinfo {author} {\bibfnamefont
  {J.~D.}\ \bibnamefont {Song}}, \bibinfo {author} {\bibfnamefont
  {S.}~\bibnamefont {Stobbe}}, \ and\ \bibinfo {author} {\bibfnamefont
  {P.}~\bibnamefont {Lodahl}},\ }\href {\doibase
  10.1103/PhysRevLett.113.093603} {\bibfield  {journal} {\bibinfo  {journal}
  {Physical Review Letters}\ }\textbf {\bibinfo {volume} {113}},\ \bibinfo
  {pages} {093603} (\bibinfo {year} {2014})}\BibitemShut {NoStop}%
\bibitem [{\citenamefont {Reithmaier}\ \emph {et~al.}(2015)\citenamefont
  {Reithmaier}, \citenamefont {Kaniber}, \citenamefont {Flassig}, \citenamefont
  {Lichtmannecker}, \citenamefont {M{\"{u}}ller}, \citenamefont {Andrejew},
  \citenamefont {Vu{\v{c}}kovi{\'{c}}}, \citenamefont {Gross},\ and\
  \citenamefont {Finley}}]{Reithmaier2015}%
  \BibitemOpen
  \bibfield  {author} {\bibinfo {author} {\bibfnamefont {G.}~\bibnamefont
  {Reithmaier}}, \bibinfo {author} {\bibfnamefont {M.}~\bibnamefont {Kaniber}},
  \bibinfo {author} {\bibfnamefont {F.}~\bibnamefont {Flassig}}, \bibinfo
  {author} {\bibfnamefont {S.}~\bibnamefont {Lichtmannecker}}, \bibinfo
  {author} {\bibfnamefont {K.}~\bibnamefont {M{\"{u}}ller}}, \bibinfo {author}
  {\bibfnamefont {A.}~\bibnamefont {Andrejew}}, \bibinfo {author}
  {\bibfnamefont {J.}~\bibnamefont {Vu{\v{c}}kovi{\'{c}}}}, \bibinfo {author}
  {\bibfnamefont {R.}~\bibnamefont {Gross}}, \ and\ \bibinfo {author}
  {\bibfnamefont {J.~J.}\ \bibnamefont {Finley}},\ }\href {\doibase
  10.1021/acs.nanolett.5b01444} {\bibfield  {journal} {\bibinfo  {journal}
  {Nano Letters}\ }\textbf {\bibinfo {volume} {15}},\ \bibinfo {pages} {5208}
  (\bibinfo {year} {2015})}\BibitemShut {NoStop}%
\bibitem [{\citenamefont {Davanco}\ \emph {et~al.}(2017)\citenamefont
  {Davanco}, \citenamefont {Liu}, \citenamefont {Sapienza}, \citenamefont
  {Zhang}, \citenamefont {{De Miranda Cardoso}}, \citenamefont {Verma},
  \citenamefont {Mirin}, \citenamefont {Nam}, \citenamefont {Liu},\ and\
  \citenamefont {Srinivasan}}]{Davanco2017}%
  \BibitemOpen
  \bibfield  {author} {\bibinfo {author} {\bibfnamefont {M.}~\bibnamefont
  {Davanco}}, \bibinfo {author} {\bibfnamefont {J.}~\bibnamefont {Liu}},
  \bibinfo {author} {\bibfnamefont {L.}~\bibnamefont {Sapienza}}, \bibinfo
  {author} {\bibfnamefont {C.-Z.}\ \bibnamefont {Zhang}}, \bibinfo {author}
  {\bibfnamefont {J.~V.}\ \bibnamefont {{De Miranda Cardoso}}}, \bibinfo
  {author} {\bibfnamefont {V.}~\bibnamefont {Verma}}, \bibinfo {author}
  {\bibfnamefont {R.}~\bibnamefont {Mirin}}, \bibinfo {author} {\bibfnamefont
  {S.~W.}\ \bibnamefont {Nam}}, \bibinfo {author} {\bibfnamefont
  {L.}~\bibnamefont {Liu}}, \ and\ \bibinfo {author} {\bibfnamefont
  {K.}~\bibnamefont {Srinivasan}},\ }\href {\doibase
  10.1038/s41467-017-00987-6} {\bibfield  {journal} {\bibinfo  {journal}
  {Nature Communications}\ }\textbf {\bibinfo {volume} {8}},\ \bibinfo {pages}
  {889} (\bibinfo {year} {2017})}\BibitemShut {NoStop}%
\bibitem [{\citenamefont {Elshaari}\ \emph {et~al.}(2017)\citenamefont
  {Elshaari}, \citenamefont {Zadeh}, \citenamefont {Fognini}, \citenamefont
  {Reimer}, \citenamefont {Dalacu}, \citenamefont {Poole}, \citenamefont
  {Zwiller},\ and\ \citenamefont {J{\"{o}}ns}}]{Elshaari2017}%
  \BibitemOpen
  \bibfield  {author} {\bibinfo {author} {\bibfnamefont {A.~W.}\ \bibnamefont
  {Elshaari}}, \bibinfo {author} {\bibfnamefont {I.~E.}\ \bibnamefont {Zadeh}},
  \bibinfo {author} {\bibfnamefont {A.}~\bibnamefont {Fognini}}, \bibinfo
  {author} {\bibfnamefont {M.~E.}\ \bibnamefont {Reimer}}, \bibinfo {author}
  {\bibfnamefont {D.}~\bibnamefont {Dalacu}}, \bibinfo {author} {\bibfnamefont
  {P.~J.}\ \bibnamefont {Poole}}, \bibinfo {author} {\bibfnamefont
  {V.}~\bibnamefont {Zwiller}}, \ and\ \bibinfo {author} {\bibfnamefont
  {K.~D.}\ \bibnamefont {J{\"{o}}ns}},\ }\href {\doibase
  10.1038/s41467-017-00486-8} {\bibfield  {journal} {\bibinfo  {journal}
  {Nature Communications}\ }\textbf {\bibinfo {volume} {8}},\ \bibinfo {pages}
  {379} (\bibinfo {year} {2017})}\BibitemShut {NoStop}%
\bibitem [{\citenamefont {Kim}\ \emph {et~al.}(2017)\citenamefont {Kim},
  \citenamefont {Aghaeimeibodi}, \citenamefont {Richardson}, \citenamefont
  {Leavitt}, \citenamefont {Englund},\ and\ \citenamefont {Waks}}]{Kim2017}%
  \BibitemOpen
  \bibfield  {author} {\bibinfo {author} {\bibfnamefont {J.-H.}\ \bibnamefont
  {Kim}}, \bibinfo {author} {\bibfnamefont {S.}~\bibnamefont {Aghaeimeibodi}},
  \bibinfo {author} {\bibfnamefont {C.~J.~K.}\ \bibnamefont {Richardson}},
  \bibinfo {author} {\bibfnamefont {R.~P.}\ \bibnamefont {Leavitt}}, \bibinfo
  {author} {\bibfnamefont {D.}~\bibnamefont {Englund}}, \ and\ \bibinfo
  {author} {\bibfnamefont {E.}~\bibnamefont {Waks}},\ }\href {\doibase
  10.1021/acs.nanolett.7b03220} {\bibfield  {journal} {\bibinfo  {journal}
  {Nano Letters}\ }\textbf {\bibinfo {volume} {17}},\ \bibinfo {pages} {7394}
  (\bibinfo {year} {2017})}\BibitemShut {NoStop}%
\bibitem [{\citenamefont {Midolo}\ \emph {et~al.}(2017)\citenamefont {Midolo},
  \citenamefont {Hansen}, \citenamefont {Zhang}, \citenamefont {Papon},
  \citenamefont {Schott}, \citenamefont {Ludwig}, \citenamefont {Wieck},
  \citenamefont {Lodahl},\ and\ \citenamefont {Stobbe}}]{Midolo2017}%
  \BibitemOpen
  \bibfield  {author} {\bibinfo {author} {\bibfnamefont {L.}~\bibnamefont
  {Midolo}}, \bibinfo {author} {\bibfnamefont {S.~L.}\ \bibnamefont {Hansen}},
  \bibinfo {author} {\bibfnamefont {W.}~\bibnamefont {Zhang}}, \bibinfo
  {author} {\bibfnamefont {C.}~\bibnamefont {Papon}}, \bibinfo {author}
  {\bibfnamefont {R.}~\bibnamefont {Schott}}, \bibinfo {author} {\bibfnamefont
  {A.}~\bibnamefont {Ludwig}}, \bibinfo {author} {\bibfnamefont {A.~D.}\
  \bibnamefont {Wieck}}, \bibinfo {author} {\bibfnamefont {P.}~\bibnamefont
  {Lodahl}}, \ and\ \bibinfo {author} {\bibfnamefont {S.}~\bibnamefont
  {Stobbe}},\ }\href {\doibase 10.1364/OE.25.033514} {\bibfield  {journal}
  {\bibinfo  {journal} {Opt. Express}\ }\textbf {\bibinfo {volume} {25}},\
  \bibinfo {pages} {33514} (\bibinfo {year} {2017})}\BibitemShut {NoStop}%
\bibitem [{\citenamefont {Wang}\ \emph {et~al.}(2014)\citenamefont {Wang},
  \citenamefont {Santamato}, \citenamefont {Jiang}, \citenamefont {Bonneau},
  \citenamefont {Engin}, \citenamefont {Silverstone}, \citenamefont {Lermer},
  \citenamefont {Beetz}, \citenamefont {Kamp}, \citenamefont {H{\"{o}}fling},
  \citenamefont {Tanner}, \citenamefont {Natarajan}, \citenamefont {Hadfield},
  \citenamefont {Dorenbos}, \citenamefont {Zwiller}, \citenamefont {O'Brien},\
  and\ \citenamefont {Thompson}}]{Wang2014a}%
  \BibitemOpen
  \bibfield  {author} {\bibinfo {author} {\bibfnamefont {J.}~\bibnamefont
  {Wang}}, \bibinfo {author} {\bibfnamefont {A.}~\bibnamefont {Santamato}},
  \bibinfo {author} {\bibfnamefont {P.}~\bibnamefont {Jiang}}, \bibinfo
  {author} {\bibfnamefont {D.}~\bibnamefont {Bonneau}}, \bibinfo {author}
  {\bibfnamefont {E.}~\bibnamefont {Engin}}, \bibinfo {author} {\bibfnamefont
  {J.~W.}\ \bibnamefont {Silverstone}}, \bibinfo {author} {\bibfnamefont
  {M.}~\bibnamefont {Lermer}}, \bibinfo {author} {\bibfnamefont
  {J.}~\bibnamefont {Beetz}}, \bibinfo {author} {\bibfnamefont
  {M.}~\bibnamefont {Kamp}}, \bibinfo {author} {\bibfnamefont {S.}~\bibnamefont
  {H{\"{o}}fling}}, \bibinfo {author} {\bibfnamefont {M.~G.}\ \bibnamefont
  {Tanner}}, \bibinfo {author} {\bibfnamefont {C.~M.}\ \bibnamefont
  {Natarajan}}, \bibinfo {author} {\bibfnamefont {R.~H.}\ \bibnamefont
  {Hadfield}}, \bibinfo {author} {\bibfnamefont {S.~N.}\ \bibnamefont
  {Dorenbos}}, \bibinfo {author} {\bibfnamefont {V.}~\bibnamefont {Zwiller}},
  \bibinfo {author} {\bibfnamefont {J.~L.}\ \bibnamefont {O'Brien}}, \ and\
  \bibinfo {author} {\bibfnamefont {M.~G.}\ \bibnamefont {Thompson}},\ }\href
  {\doibase 10.1016/j.optcom.2014.02.040} {\bibfield  {journal} {\bibinfo
  {journal} {Optics Communications}\ }\textbf {\bibinfo {volume} {327}},\
  \bibinfo {pages} {49} (\bibinfo {year} {2014})}\BibitemShut {NoStop}%
\bibitem [{\citenamefont {Prtljaga}\ \emph {et~al.}(2014)\citenamefont
  {Prtljaga}, \citenamefont {Coles}, \citenamefont {O'Hara}, \citenamefont
  {Royall}, \citenamefont {Clarke}, \citenamefont {Fox},\ and\ \citenamefont
  {Skolnick}}]{Prtljaga2014}%
  \BibitemOpen
  \bibfield  {author} {\bibinfo {author} {\bibfnamefont {N.}~\bibnamefont
  {Prtljaga}}, \bibinfo {author} {\bibfnamefont {R.~J.}\ \bibnamefont {Coles}},
  \bibinfo {author} {\bibfnamefont {J.}~\bibnamefont {O'Hara}}, \bibinfo
  {author} {\bibfnamefont {B.}~\bibnamefont {Royall}}, \bibinfo {author}
  {\bibfnamefont {E.}~\bibnamefont {Clarke}}, \bibinfo {author} {\bibfnamefont
  {A.~M.}\ \bibnamefont {Fox}}, \ and\ \bibinfo {author} {\bibfnamefont
  {M.~S.}\ \bibnamefont {Skolnick}},\ }\href {\doibase 10.1063/1.4883374}
  {\bibfield  {journal} {\bibinfo  {journal} {Applied Physics Letters}\
  }\textbf {\bibinfo {volume} {104}},\ \bibinfo {pages} {231107} (\bibinfo
  {year} {2014})}\BibitemShut {NoStop}%
\bibitem [{\citenamefont {Harris}\ \emph {et~al.}(2014)\citenamefont {Harris},
  \citenamefont {Grassani}, \citenamefont {Simbula}, \citenamefont {Pant},
  \citenamefont {Galli}, \citenamefont {Baehr-Jones}, \citenamefont {Hochberg},
  \citenamefont {Englund}, \citenamefont {Bajoni},\ and\ \citenamefont
  {Galland}}]{Harris2014}%
  \BibitemOpen
  \bibfield  {author} {\bibinfo {author} {\bibfnamefont {N.~C.}\ \bibnamefont
  {Harris}}, \bibinfo {author} {\bibfnamefont {D.}~\bibnamefont {Grassani}},
  \bibinfo {author} {\bibfnamefont {A.}~\bibnamefont {Simbula}}, \bibinfo
  {author} {\bibfnamefont {M.}~\bibnamefont {Pant}}, \bibinfo {author}
  {\bibfnamefont {M.}~\bibnamefont {Galli}}, \bibinfo {author} {\bibfnamefont
  {T.}~\bibnamefont {Baehr-Jones}}, \bibinfo {author} {\bibfnamefont
  {M.}~\bibnamefont {Hochberg}}, \bibinfo {author} {\bibfnamefont
  {D.}~\bibnamefont {Englund}}, \bibinfo {author} {\bibfnamefont
  {D.}~\bibnamefont {Bajoni}}, \ and\ \bibinfo {author} {\bibfnamefont
  {C.}~\bibnamefont {Galland}},\ }\href {\doibase 10.1103/PhysRevX.4.041047}
  {\bibfield  {journal} {\bibinfo  {journal} {Physical Review X}\ }\textbf
  {\bibinfo {volume} {4}},\ \bibinfo {pages} {041047} (\bibinfo {year}
  {2014})}\BibitemShut {NoStop}%
\bibitem [{\citenamefont {Kaniber}\ \emph {et~al.}(2016)\citenamefont
  {Kaniber}, \citenamefont {Flassig}, \citenamefont {Reithmaier}, \citenamefont
  {Gross},\ and\ \citenamefont {Finley}}]{Kaniber2016}%
  \BibitemOpen
  \bibfield  {author} {\bibinfo {author} {\bibfnamefont {M.}~\bibnamefont
  {Kaniber}}, \bibinfo {author} {\bibfnamefont {F.}~\bibnamefont {Flassig}},
  \bibinfo {author} {\bibfnamefont {G.}~\bibnamefont {Reithmaier}}, \bibinfo
  {author} {\bibfnamefont {R.}~\bibnamefont {Gross}}, \ and\ \bibinfo {author}
  {\bibfnamefont {J.~J.}\ \bibnamefont {Finley}},\ }\href {\doibase
  10.1007/s00340-016-6376-1} {\bibfield  {journal} {\bibinfo  {journal}
  {Applied Physics B}\ }\textbf {\bibinfo {volume} {122}},\ \bibinfo {pages}
  {115} (\bibinfo {year} {2016})}\BibitemShut {NoStop}%
\bibitem [{\citenamefont {Lund-Hansen}\ \emph {et~al.}(2008)\citenamefont
  {Lund-Hansen}, \citenamefont {Stobbe}, \citenamefont {Julsgaard},
  \citenamefont {Thyrrestrup}, \citenamefont {S{\"{u}}nner}, \citenamefont
  {Kamp}, \citenamefont {Forchel},\ and\ \citenamefont
  {Lodahl}}]{Lund-Hansen2008}%
  \BibitemOpen
  \bibfield  {author} {\bibinfo {author} {\bibfnamefont {T.}~\bibnamefont
  {Lund-Hansen}}, \bibinfo {author} {\bibfnamefont {S.}~\bibnamefont {Stobbe}},
  \bibinfo {author} {\bibfnamefont {B.}~\bibnamefont {Julsgaard}}, \bibinfo
  {author} {\bibfnamefont {H.}~\bibnamefont {Thyrrestrup}}, \bibinfo {author}
  {\bibfnamefont {T.}~\bibnamefont {S{\"{u}}nner}}, \bibinfo {author}
  {\bibfnamefont {M.}~\bibnamefont {Kamp}}, \bibinfo {author} {\bibfnamefont
  {A.}~\bibnamefont {Forchel}}, \ and\ \bibinfo {author} {\bibfnamefont
  {P.}~\bibnamefont {Lodahl}},\ }\href {\doibase
  10.1103/PhysRevLett.101.113903} {\bibfield  {journal} {\bibinfo  {journal}
  {Physical Review Letters}\ }\textbf {\bibinfo {volume} {101}},\ \bibinfo
  {pages} {113903} (\bibinfo {year} {2008})}\BibitemShut {NoStop}%
\bibitem [{\citenamefont {Stepanov}\ \emph {et~al.}(2015)\citenamefont
  {Stepanov}, \citenamefont {Delga}, \citenamefont {Zang}, \citenamefont
  {Bleuse}, \citenamefont {Dupuy}, \citenamefont {Peinke}, \citenamefont
  {Lalanne}, \citenamefont {G{\'{e}}rard},\ and\ \citenamefont
  {Claudon}}]{Stepanov2015}%
  \BibitemOpen
  \bibfield  {author} {\bibinfo {author} {\bibfnamefont {P.}~\bibnamefont
  {Stepanov}}, \bibinfo {author} {\bibfnamefont {A.}~\bibnamefont {Delga}},
  \bibinfo {author} {\bibfnamefont {X.}~\bibnamefont {Zang}}, \bibinfo {author}
  {\bibfnamefont {J.}~\bibnamefont {Bleuse}}, \bibinfo {author} {\bibfnamefont
  {E.}~\bibnamefont {Dupuy}}, \bibinfo {author} {\bibfnamefont
  {E.}~\bibnamefont {Peinke}}, \bibinfo {author} {\bibfnamefont
  {P.}~\bibnamefont {Lalanne}}, \bibinfo {author} {\bibfnamefont {J.-M.}\
  \bibnamefont {G{\'{e}}rard}}, \ and\ \bibinfo {author} {\bibfnamefont
  {J.}~\bibnamefont {Claudon}},\ }\href {\doibase 10.1063/1.4906921} {\bibfield
   {journal} {\bibinfo  {journal} {Applied Physics Letters}\ }\textbf {\bibinfo
  {volume} {106}},\ \bibinfo {pages} {041112} (\bibinfo {year}
  {2015})}\BibitemShut {NoStop}%
\bibitem [{\citenamefont {Makhonin}\ \emph {et~al.}(2014)\citenamefont
  {Makhonin}, \citenamefont {Dixon}, \citenamefont {Coles}, \citenamefont
  {Royall}, \citenamefont {Luxmoore}, \citenamefont {Clarke}, \citenamefont
  {Hugues}, \citenamefont {Skolnick},\ and\ \citenamefont
  {Fox}}]{Makhonin2014}%
  \BibitemOpen
  \bibfield  {author} {\bibinfo {author} {\bibfnamefont {M.~N.}\ \bibnamefont
  {Makhonin}}, \bibinfo {author} {\bibfnamefont {J.~E.}\ \bibnamefont {Dixon}},
  \bibinfo {author} {\bibfnamefont {R.~J.}\ \bibnamefont {Coles}}, \bibinfo
  {author} {\bibfnamefont {B.}~\bibnamefont {Royall}}, \bibinfo {author}
  {\bibfnamefont {I.~J.}\ \bibnamefont {Luxmoore}}, \bibinfo {author}
  {\bibfnamefont {E.}~\bibnamefont {Clarke}}, \bibinfo {author} {\bibfnamefont
  {M.}~\bibnamefont {Hugues}}, \bibinfo {author} {\bibfnamefont {M.~S.}\
  \bibnamefont {Skolnick}}, \ and\ \bibinfo {author} {\bibfnamefont {A.~M.}\
  \bibnamefont {Fox}},\ }\href {\doibase 10.1021/nl5032937} {\bibfield
  {journal} {\bibinfo  {journal} {Nano Letters}\ }\textbf {\bibinfo {volume}
  {14}},\ \bibinfo {pages} {6997} (\bibinfo {year} {2014})}\BibitemShut
  {NoStop}%
\bibitem [{\citenamefont {Kalliakos}\ \emph {et~al.}(2016)\citenamefont
  {Kalliakos}, \citenamefont {Brody}, \citenamefont {Bennett}, \citenamefont
  {Ellis}, \citenamefont {Skiba-Szymanska}, \citenamefont {Farrer},
  \citenamefont {Griffiths}, \citenamefont {Ritchie},\ and\ \citenamefont
  {Shields}}]{Kalliakos2016}%
  \BibitemOpen
  \bibfield  {author} {\bibinfo {author} {\bibfnamefont {S.}~\bibnamefont
  {Kalliakos}}, \bibinfo {author} {\bibfnamefont {Y.}~\bibnamefont {Brody}},
  \bibinfo {author} {\bibfnamefont {A.~J.}\ \bibnamefont {Bennett}}, \bibinfo
  {author} {\bibfnamefont {D.~J.~P.}\ \bibnamefont {Ellis}}, \bibinfo {author}
  {\bibfnamefont {J.}~\bibnamefont {Skiba-Szymanska}}, \bibinfo {author}
  {\bibfnamefont {I.}~\bibnamefont {Farrer}}, \bibinfo {author} {\bibfnamefont
  {J.~P.}\ \bibnamefont {Griffiths}}, \bibinfo {author} {\bibfnamefont {D.~A.}\
  \bibnamefont {Ritchie}}, \ and\ \bibinfo {author} {\bibfnamefont {A.~J.}\
  \bibnamefont {Shields}},\ }\href {\doibase 10.1063/1.4964888} {\bibfield
  {journal} {\bibinfo  {journal} {Applied Physics Letters}\ }\textbf {\bibinfo
  {volume} {109}},\ \bibinfo {pages} {151112} (\bibinfo {year}
  {2016})}\BibitemShut {NoStop}%
\bibitem [{\citenamefont {Schwartz}\ \emph {et~al.}(2016)\citenamefont
  {Schwartz}, \citenamefont {Rengstl}, \citenamefont {Herzog}, \citenamefont
  {Paul}, \citenamefont {Kettler}, \citenamefont {Portalupi}, \citenamefont
  {Jetter},\ and\ \citenamefont {Michler}}]{Schwartz2016}%
  \BibitemOpen
  \bibfield  {author} {\bibinfo {author} {\bibfnamefont {M.}~\bibnamefont
  {Schwartz}}, \bibinfo {author} {\bibfnamefont {U.}~\bibnamefont {Rengstl}},
  \bibinfo {author} {\bibfnamefont {T.}~\bibnamefont {Herzog}}, \bibinfo
  {author} {\bibfnamefont {M.}~\bibnamefont {Paul}}, \bibinfo {author}
  {\bibfnamefont {J.}~\bibnamefont {Kettler}}, \bibinfo {author} {\bibfnamefont
  {S.~L.}\ \bibnamefont {Portalupi}}, \bibinfo {author} {\bibfnamefont
  {M.}~\bibnamefont {Jetter}}, \ and\ \bibinfo {author} {\bibfnamefont
  {P.}~\bibnamefont {Michler}},\ }\href {\doibase 10.1364/OE.24.003089}
  {\bibfield  {journal} {\bibinfo  {journal} {Optics Express}\ }\textbf
  {\bibinfo {volume} {24}},\ \bibinfo {pages} {3089} (\bibinfo {year}
  {2016})}\BibitemShut {NoStop}%
\bibitem [{\citenamefont {Liu}\ \emph {et~al.}(2018)\citenamefont {Liu},
  \citenamefont {Brash}, \citenamefont {O'Hara}, \citenamefont {Martins},
  \citenamefont {Phillips}, \citenamefont {Coles}, \citenamefont {Royall},
  \citenamefont {Clarke}, \citenamefont {Bentham}, \citenamefont {Prtljaga},
  \citenamefont {Itskevich}, \citenamefont {Wilson}, \citenamefont {Skolnick},\
  and\ \citenamefont {Fox}}]{Liu2018}%
  \BibitemOpen
  \bibfield  {author} {\bibinfo {author} {\bibfnamefont {F.}~\bibnamefont
  {Liu}}, \bibinfo {author} {\bibfnamefont {A.~J.}\ \bibnamefont {Brash}},
  \bibinfo {author} {\bibfnamefont {J.}~\bibnamefont {O'Hara}}, \bibinfo
  {author} {\bibfnamefont {L.~M. P.~P.}\ \bibnamefont {Martins}}, \bibinfo
  {author} {\bibfnamefont {C.~L.}\ \bibnamefont {Phillips}}, \bibinfo {author}
  {\bibfnamefont {R.~J.}\ \bibnamefont {Coles}}, \bibinfo {author}
  {\bibfnamefont {B.}~\bibnamefont {Royall}}, \bibinfo {author} {\bibfnamefont
  {E.}~\bibnamefont {Clarke}}, \bibinfo {author} {\bibfnamefont
  {C.}~\bibnamefont {Bentham}}, \bibinfo {author} {\bibfnamefont
  {N.}~\bibnamefont {Prtljaga}}, \bibinfo {author} {\bibfnamefont {I.~E.}\
  \bibnamefont {Itskevich}}, \bibinfo {author} {\bibfnamefont {L.~R.}\
  \bibnamefont {Wilson}}, \bibinfo {author} {\bibfnamefont {M.~S.}\
  \bibnamefont {Skolnick}}, \ and\ \bibinfo {author} {\bibfnamefont {A.~M.}\
  \bibnamefont {Fox}},\ }\href {\doibase 10.1038/s41565-018-0188-x} {\bibfield
  {journal} {\bibinfo  {journal} {Nature Nanotechnology}\ }\textbf {\bibinfo
  {volume} {13}},\ \bibinfo {pages} {835} (\bibinfo {year} {2018})}\BibitemShut
  {NoStop}%
\bibitem [{\citenamefont {Pan}\ \emph {et~al.}(2012)\citenamefont {Pan},
  \citenamefont {Chen}, \citenamefont {Lu}, \citenamefont {Weinfurter},
  \citenamefont {Zeilinger},\ and\ \citenamefont {{\.{Z}}ukowski}}]{Pan2012}%
  \BibitemOpen
  \bibfield  {author} {\bibinfo {author} {\bibfnamefont {J.-W.}\ \bibnamefont
  {Pan}}, \bibinfo {author} {\bibfnamefont {Z.-B.}\ \bibnamefont {Chen}},
  \bibinfo {author} {\bibfnamefont {C.-Y.}\ \bibnamefont {Lu}}, \bibinfo
  {author} {\bibfnamefont {H.}~\bibnamefont {Weinfurter}}, \bibinfo {author}
  {\bibfnamefont {A.}~\bibnamefont {Zeilinger}}, \ and\ \bibinfo {author}
  {\bibfnamefont {M.}~\bibnamefont {{\.{Z}}ukowski}},\ }\href {\doibase
  10.1103/RevModPhys.84.777} {\bibfield  {journal} {\bibinfo  {journal}
  {Reviews of Modern Physics}\ }\textbf {\bibinfo {volume} {84}},\ \bibinfo
  {pages} {777} (\bibinfo {year} {2012})}\BibitemShut {NoStop}%
\bibitem [{\citenamefont {Press}\ \emph {et~al.}(2010)\citenamefont {Press},
  \citenamefont {{De Greve}}, \citenamefont {McMahon}, \citenamefont {Ladd},
  \citenamefont {Friess}, \citenamefont {Schneider}, \citenamefont {Kamp},
  \citenamefont {H{\"{o}}fling}, \citenamefont {Forchel},\ and\ \citenamefont
  {Yamamoto}}]{Press2010}%
  \BibitemOpen
  \bibfield  {author} {\bibinfo {author} {\bibfnamefont {D.}~\bibnamefont
  {Press}}, \bibinfo {author} {\bibfnamefont {K.}~\bibnamefont {{De Greve}}},
  \bibinfo {author} {\bibfnamefont {P.~L.}\ \bibnamefont {McMahon}}, \bibinfo
  {author} {\bibfnamefont {T.~D.}\ \bibnamefont {Ladd}}, \bibinfo {author}
  {\bibfnamefont {B.}~\bibnamefont {Friess}}, \bibinfo {author} {\bibfnamefont
  {C.}~\bibnamefont {Schneider}}, \bibinfo {author} {\bibfnamefont
  {M.}~\bibnamefont {Kamp}}, \bibinfo {author} {\bibfnamefont {S.}~\bibnamefont
  {H{\"{o}}fling}}, \bibinfo {author} {\bibfnamefont {A.}~\bibnamefont
  {Forchel}}, \ and\ \bibinfo {author} {\bibfnamefont {Y.}~\bibnamefont
  {Yamamoto}},\ }\href {\doibase 10.1038/nphoton.2010.83} {\bibfield  {journal}
  {\bibinfo  {journal} {Nature Photonics}\ }\textbf {\bibinfo {volume} {4}},\
  \bibinfo {pages} {367} (\bibinfo {year} {2010})}\BibitemShut {NoStop}%
\bibitem [{\citenamefont {Majumdar}\ \emph {et~al.}(2011)\citenamefont
  {Majumdar}, \citenamefont {Kim},\ and\ \citenamefont
  {Vu{\v{c}}kovi{\'{c}}}}]{Majumdar2011}%
  \BibitemOpen
  \bibfield  {author} {\bibinfo {author} {\bibfnamefont {A.}~\bibnamefont
  {Majumdar}}, \bibinfo {author} {\bibfnamefont {E.~D.}\ \bibnamefont {Kim}}, \
  and\ \bibinfo {author} {\bibfnamefont {J.}~\bibnamefont
  {Vu{\v{c}}kovi{\'{c}}}},\ }\href {\doibase 10.1103/PhysRevB.84.195304}
  {\bibfield  {journal} {\bibinfo  {journal} {Physical Review B}\ }\textbf
  {\bibinfo {volume} {84}},\ \bibinfo {pages} {195304} (\bibinfo {year}
  {2011})}\BibitemShut {NoStop}%
\bibitem [{\citenamefont {Iles-Smith}\ \emph {et~al.}(2017)\citenamefont
  {Iles-Smith}, \citenamefont {McCutcheon}, \citenamefont {Nazir},\ and\
  \citenamefont {M{\o}rk}}]{Iles-Smith2017}%
  \BibitemOpen
  \bibfield  {author} {\bibinfo {author} {\bibfnamefont {J.}~\bibnamefont
  {Iles-Smith}}, \bibinfo {author} {\bibfnamefont {D.~P.~S.}\ \bibnamefont
  {McCutcheon}}, \bibinfo {author} {\bibfnamefont {A.}~\bibnamefont {Nazir}}, \
  and\ \bibinfo {author} {\bibfnamefont {J.}~\bibnamefont {M{\o}rk}},\ }\href
  {\doibase 10.1038/nphoton.2017.101} {\bibfield  {journal} {\bibinfo
  {journal} {Nature Photonics}\ }\textbf {\bibinfo {volume} {11}},\ \bibinfo
  {pages} {521} (\bibinfo {year} {2017})},\ \Eprint
  {http://arxiv.org/abs/1612.04173} {1612.04173} \BibitemShut {NoStop}%
\bibitem [{\citenamefont {F{\"{o}}rstner}\ \emph {et~al.}(2003)\citenamefont
  {F{\"{o}}rstner}, \citenamefont {Weber}, \citenamefont {Danckwerts},\ and\
  \citenamefont {Knorr}}]{Forstner2003}%
  \BibitemOpen
  \bibfield  {author} {\bibinfo {author} {\bibfnamefont {J.}~\bibnamefont
  {F{\"{o}}rstner}}, \bibinfo {author} {\bibfnamefont {C.}~\bibnamefont
  {Weber}}, \bibinfo {author} {\bibfnamefont {J.}~\bibnamefont {Danckwerts}}, \
  and\ \bibinfo {author} {\bibfnamefont {A.}~\bibnamefont {Knorr}},\ }\href
  {\doibase 10.1103/PhysRevLett.91.127401} {\bibfield  {journal} {\bibinfo
  {journal} {Physical Review Letters}\ }\textbf {\bibinfo {volume} {91}},\
  \bibinfo {pages} {127401} (\bibinfo {year} {2003})}\BibitemShut {NoStop}%
\bibitem [{\citenamefont {Hanschke}\ \emph {et~al.}(2018)\citenamefont
  {Hanschke}, \citenamefont {Fischer}, \citenamefont {Appel}, \citenamefont
  {Lukin}, \citenamefont {Wierzbowski}, \citenamefont {Sun}, \citenamefont
  {Trivedi}, \citenamefont {Vu{\v{c}}kovi{\'{c}}}, \citenamefont {Finley},\
  and\ \citenamefont {M{\"{u}}ller}}]{Hanschke2018}%
  \BibitemOpen
  \bibfield  {author} {\bibinfo {author} {\bibfnamefont {L.}~\bibnamefont
  {Hanschke}}, \bibinfo {author} {\bibfnamefont {K.~A.}\ \bibnamefont
  {Fischer}}, \bibinfo {author} {\bibfnamefont {S.}~\bibnamefont {Appel}},
  \bibinfo {author} {\bibfnamefont {D.}~\bibnamefont {Lukin}}, \bibinfo
  {author} {\bibfnamefont {J.}~\bibnamefont {Wierzbowski}}, \bibinfo {author}
  {\bibfnamefont {S.}~\bibnamefont {Sun}}, \bibinfo {author} {\bibfnamefont
  {R.}~\bibnamefont {Trivedi}}, \bibinfo {author} {\bibfnamefont
  {J.}~\bibnamefont {Vu{\v{c}}kovi{\'{c}}}}, \bibinfo {author} {\bibfnamefont
  {J.~J.}\ \bibnamefont {Finley}}, \ and\ \bibinfo {author} {\bibfnamefont
  {K.}~\bibnamefont {M{\"{u}}ller}},\ }\href {\doibase
  10.1038/s41534-018-0092-0} {\bibfield  {journal} {\bibinfo  {journal} {npj
  Quantum Information}\ }\textbf {\bibinfo {volume} {4}},\ \bibinfo {pages}
  {43} (\bibinfo {year} {2018})},\ \Eprint {http://arxiv.org/abs/1801.01672}
  {1801.01672} \BibitemShut {NoStop}%
\bibitem [{\citenamefont {Fowler}\ \emph {et~al.}(2009)\citenamefont {Fowler},
  \citenamefont {Stephens},\ and\ \citenamefont {Groszkowski}}]{Fowler2009}%
  \BibitemOpen
  \bibfield  {author} {\bibinfo {author} {\bibfnamefont {A.~G.}\ \bibnamefont
  {Fowler}}, \bibinfo {author} {\bibfnamefont {A.~M.}\ \bibnamefont
  {Stephens}}, \ and\ \bibinfo {author} {\bibfnamefont {P.}~\bibnamefont
  {Groszkowski}},\ }\href {\doibase 10.1103/PhysRevA.80.052312} {\bibfield
  {journal} {\bibinfo  {journal} {Physical Review A}\ }\textbf {\bibinfo
  {volume} {80}},\ \bibinfo {pages} {052312} (\bibinfo {year}
  {2009})}\BibitemShut {NoStop}%
\bibitem [{\citenamefont {Knill}(2005)}]{Knill2005}%
  \BibitemOpen
  \bibfield  {author} {\bibinfo {author} {\bibfnamefont {E.}~\bibnamefont
  {Knill}},\ }\href {\doibase 10.1038/nature03350} {\bibfield  {journal}
  {\bibinfo  {journal} {Nature}\ }\textbf {\bibinfo {volume} {434}},\ \bibinfo
  {pages} {39} (\bibinfo {year} {2005})},\ \Eprint
  {http://arxiv.org/abs/0410199} {0410199 [quant-ph]} \BibitemShut {NoStop}%
\bibitem [{\citenamefont {Wei}\ \emph {et~al.}(2014)\citenamefont {Wei},
  \citenamefont {He}, \citenamefont {Chen}, \citenamefont {Hu}, \citenamefont
  {He}, \citenamefont {Wu}, \citenamefont {Schneider}, \citenamefont {Kamp},
  \citenamefont {H{\"{o}}fling}, \citenamefont {Lu},\ and\ \citenamefont
  {Pan}}]{Wei2014a}%
  \BibitemOpen
  \bibfield  {author} {\bibinfo {author} {\bibfnamefont {Y.-J.}\ \bibnamefont
  {Wei}}, \bibinfo {author} {\bibfnamefont {Y.-M.}\ \bibnamefont {He}},
  \bibinfo {author} {\bibfnamefont {M.-C.}\ \bibnamefont {Chen}}, \bibinfo
  {author} {\bibfnamefont {Y.-N.}\ \bibnamefont {Hu}}, \bibinfo {author}
  {\bibfnamefont {Y.}~\bibnamefont {He}}, \bibinfo {author} {\bibfnamefont
  {D.}~\bibnamefont {Wu}}, \bibinfo {author} {\bibfnamefont {C.}~\bibnamefont
  {Schneider}}, \bibinfo {author} {\bibfnamefont {M.}~\bibnamefont {Kamp}},
  \bibinfo {author} {\bibfnamefont {S.}~\bibnamefont {H{\"{o}}fling}}, \bibinfo
  {author} {\bibfnamefont {C.-Y.}\ \bibnamefont {Lu}}, \ and\ \bibinfo {author}
  {\bibfnamefont {J.-W.}\ \bibnamefont {Pan}},\ }\href {\doibase
  10.1021/nl503081n} {\bibfield  {journal} {\bibinfo  {journal} {Nano Letters}\
  }\textbf {\bibinfo {volume} {14}},\ \bibinfo {pages} {6515} (\bibinfo {year}
  {2014})}\BibitemShut {NoStop}%
\bibitem [{\citenamefont {Varnava}\ \emph {et~al.}(2008)\citenamefont
  {Varnava}, \citenamefont {Browne},\ and\ \citenamefont
  {Rudolph}}]{Varnava2008}%
  \BibitemOpen
  \bibfield  {author} {\bibinfo {author} {\bibfnamefont {M.}~\bibnamefont
  {Varnava}}, \bibinfo {author} {\bibfnamefont {D.~E.}\ \bibnamefont {Browne}},
  \ and\ \bibinfo {author} {\bibfnamefont {T.}~\bibnamefont {Rudolph}},\ }\href
  {\doibase 10.1103/PhysRevLett.100.060502} {\bibfield  {journal} {\bibinfo
  {journal} {Physical Review Letters}\ }\textbf {\bibinfo {volume} {100}},\
  \bibinfo {pages} {060502} (\bibinfo {year} {2008})}\BibitemShut {NoStop}%
\bibitem [{\citenamefont {Wang}\ \emph {et~al.}(2016)\citenamefont {Wang},
  \citenamefont {Chen}, \citenamefont {Li}, \citenamefont {Huang},
  \citenamefont {Liu}, \citenamefont {Chen}, \citenamefont {Luo}, \citenamefont
  {Su}, \citenamefont {Wu}, \citenamefont {Li}, \citenamefont {Lu},
  \citenamefont {Hu}, \citenamefont {Jiang}, \citenamefont {Peng},
  \citenamefont {Li}, \citenamefont {Liu}, \citenamefont {Chen}, \citenamefont
  {Lu},\ and\ \citenamefont {Pan}}]{Wang2016}%
  \BibitemOpen
  \bibfield  {author} {\bibinfo {author} {\bibfnamefont {X.-L.}\ \bibnamefont
  {Wang}}, \bibinfo {author} {\bibfnamefont {L.-K.}\ \bibnamefont {Chen}},
  \bibinfo {author} {\bibfnamefont {W.}~\bibnamefont {Li}}, \bibinfo {author}
  {\bibfnamefont {H.-L.}\ \bibnamefont {Huang}}, \bibinfo {author}
  {\bibfnamefont {C.}~\bibnamefont {Liu}}, \bibinfo {author} {\bibfnamefont
  {C.}~\bibnamefont {Chen}}, \bibinfo {author} {\bibfnamefont {Y.-H.}\
  \bibnamefont {Luo}}, \bibinfo {author} {\bibfnamefont {Z.-E.}\ \bibnamefont
  {Su}}, \bibinfo {author} {\bibfnamefont {D.}~\bibnamefont {Wu}}, \bibinfo
  {author} {\bibfnamefont {Z.-D.}\ \bibnamefont {Li}}, \bibinfo {author}
  {\bibfnamefont {H.}~\bibnamefont {Lu}}, \bibinfo {author} {\bibfnamefont
  {Y.}~\bibnamefont {Hu}}, \bibinfo {author} {\bibfnamefont {X.}~\bibnamefont
  {Jiang}}, \bibinfo {author} {\bibfnamefont {C.-Z.}\ \bibnamefont {Peng}},
  \bibinfo {author} {\bibfnamefont {L.}~\bibnamefont {Li}}, \bibinfo {author}
  {\bibfnamefont {N.-L.}\ \bibnamefont {Liu}}, \bibinfo {author} {\bibfnamefont
  {Y.-A.}\ \bibnamefont {Chen}}, \bibinfo {author} {\bibfnamefont {C.-Y.}\
  \bibnamefont {Lu}}, \ and\ \bibinfo {author} {\bibfnamefont {J.-W.}\
  \bibnamefont {Pan}},\ }\href {\doibase 10.1103/PhysRevLett.117.210502}
  {\bibfield  {journal} {\bibinfo  {journal} {Physical Review Letters}\
  }\textbf {\bibinfo {volume} {117}},\ \bibinfo {pages} {210502} (\bibinfo
  {year} {2016})}\BibitemShut {NoStop}%
\bibitem [{\citenamefont {Kaneda}\ and\ \citenamefont
  {Kwiat}(2018)}]{Kaneda2018}%
  \BibitemOpen
  \bibfield  {author} {\bibinfo {author} {\bibfnamefont {F.}~\bibnamefont
  {Kaneda}}\ and\ \bibinfo {author} {\bibfnamefont {P.~G.}\ \bibnamefont
  {Kwiat}},\ }\href@noop {} {\bibfield  {journal} {\bibinfo  {journal} {arXiv}\
  ,\ \bibinfo {pages} {1803.04803}} (\bibinfo {year} {2018})},\ \Eprint
  {http://arxiv.org/abs/1803.04803v1} {arXiv:1803.04803v1} \BibitemShut
  {NoStop}%
\bibitem [{\citenamefont {Carolan}\ \emph {et~al.}(2015)\citenamefont
  {Carolan}, \citenamefont {Harrold}, \citenamefont {Sparrow}, \citenamefont
  {Mart{\'{i}}n-L{\'{o}}pez}, \citenamefont {Russell}, \citenamefont
  {Silverstone}, \citenamefont {Shadbolt}, \citenamefont {Matsuda},
  \citenamefont {Oguma}, \citenamefont {Itoh}, \citenamefont {Marshall},
  \citenamefont {Thompson}, \citenamefont {Matthews}, \citenamefont
  {Hashimoto}, \citenamefont {O'Brien},\ and\ \citenamefont
  {Laing}}]{Carolan2015}%
  \BibitemOpen
  \bibfield  {author} {\bibinfo {author} {\bibfnamefont {J.}~\bibnamefont
  {Carolan}}, \bibinfo {author} {\bibfnamefont {C.}~\bibnamefont {Harrold}},
  \bibinfo {author} {\bibfnamefont {C.}~\bibnamefont {Sparrow}}, \bibinfo
  {author} {\bibfnamefont {E.}~\bibnamefont {Mart{\'{i}}n-L{\'{o}}pez}},
  \bibinfo {author} {\bibfnamefont {N.~J.}\ \bibnamefont {Russell}}, \bibinfo
  {author} {\bibfnamefont {J.~W.}\ \bibnamefont {Silverstone}}, \bibinfo
  {author} {\bibfnamefont {P.~J.}\ \bibnamefont {Shadbolt}}, \bibinfo {author}
  {\bibfnamefont {N.}~\bibnamefont {Matsuda}}, \bibinfo {author} {\bibfnamefont
  {M.}~\bibnamefont {Oguma}}, \bibinfo {author} {\bibfnamefont
  {M.}~\bibnamefont {Itoh}}, \bibinfo {author} {\bibfnamefont {G.~D.}\
  \bibnamefont {Marshall}}, \bibinfo {author} {\bibfnamefont {M.~G.}\
  \bibnamefont {Thompson}}, \bibinfo {author} {\bibfnamefont {J.~C.~F.}\
  \bibnamefont {Matthews}}, \bibinfo {author} {\bibfnamefont {T.}~\bibnamefont
  {Hashimoto}}, \bibinfo {author} {\bibfnamefont {J.~L.}\ \bibnamefont
  {O'Brien}}, \ and\ \bibinfo {author} {\bibfnamefont {A.}~\bibnamefont
  {Laing}},\ }\href {\doibase 10.1126/science.aab3642} {\bibfield  {journal}
  {\bibinfo  {journal} {Science}\ }\textbf {\bibinfo {volume} {349}},\ \bibinfo
  {pages} {711} (\bibinfo {year} {2015})}\BibitemShut {NoStop}%
\end{thebibliography}%

\begin{widetext}

\section{Supplemental Material}

\renewcommand\thefigure{S\arabic{figure}} 
\setcounter{figure}{0}  

\section{Experimental details}
\textbf{Sample structure.} To fabricate our integrated single photon source waveguide device we use a semiconductor sample which contains self-assembled In(Ga)As QDs grown by the Stranski-Krastanow method at the center of a planar GaAs microcavity. The lower and upper cavity mirrors contain 24 and 5 pairs of Al$_{0.9}$Ga$_{0.1}$As/GaAs $\lambda$/4-layers, respectively, yielding a quality factor of $\sim$200. A $\delta$-doping layer of Si donors with a surface density of roughly $\sim$10$^{10}$~cm$^{-2}$ was grown 10~nm below the layer of QDs to probabilistically dope them. The full layer structure is shown in Figure~\ref{fig:sample}. In order to fabricate ridge waveguides, the top mirror layer along with GaAs cavity is etched down, forming the ridge with a width of $\sim$2~$\mu$m or $\sim$0.8~$\mu$m and a height of $\sim$1.25~$\mu$m. Ridges have been defined by e-beam lithography and reactive ion etching followed by sidewalls passivation with $\sim$10~nm of SiN. After processing, the sample was cleaved perpendicularly to the WGs in order to get clear side access to the ridge facets. 

\begin{figure}[h]
	\includegraphics[width=5in]{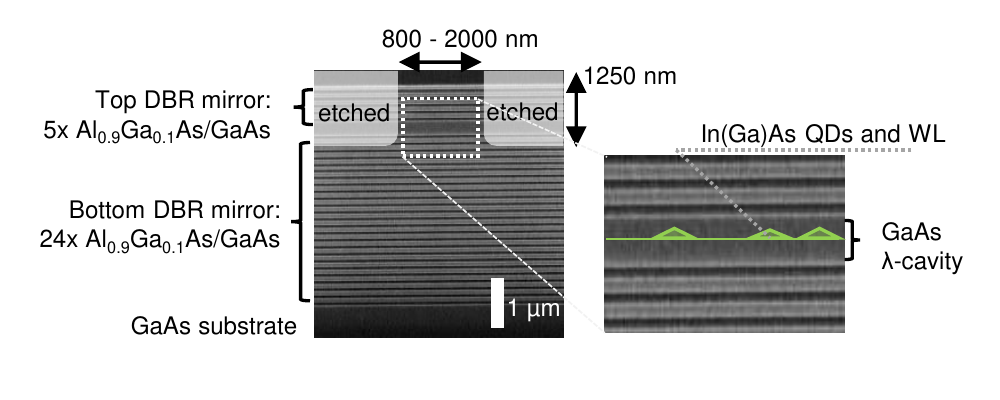}
	\caption{\label{fig:sample}Sample structure. Planar sample SEM cross-section image with visible layers and schematically marked areas for etching. Quantum dot layer is placed inside $\lambda$ cavity sandwiched between two distributed Bragg Reflectors consisting of the 5/24 alternating $\lambda$/4-thick layers of Al$_{0.2}$Ga$_{0.8}$As and GaAs.  
	}
\end{figure}

\textbf{Optical set-up.} For all experiments, the sample is kept in a low-vibrations closed-cycle cryostat (attoDry800) at temperatures of $\sim$4.5~K. The cryostat is equipped with two optical windows allowing for access from both side and top of the sample. A spectroscopic setup consisting of two independent perpendicularly aligned optical paths is employed as shown schematically in Figure~\ref{fig:setup}. QDs embedded into WG are excited from the top through a first microscope objective with NA~=~0.4, while the emission signal is detected from a side facet of the WG with a second objective with NA~=~0.4. Photoluminescence (PL) and resonance fluorescence signals are then passed through spatial filter and polarization optics. For light polarization control, a half-wave plate combined with a linear polarizer is used in both excitation and detection paths. The collected light is analyzed by a high-resolution monochromator equipped with a liquid nitrogen-cooled low-noise charge coupled device detector, featuring a spectral resolution of $\sim$20~$\mu$eV. For non-resonant PL experiments, a 660~nm continuous wave laser is used while for resonance fluorescence investigations a tunable mode-locked Ti:Sapphire laser with repetition rate of 82~MHz and pulse width of around 2~ps is used. 

\begin{figure}
	\includegraphics[width=4.5in]{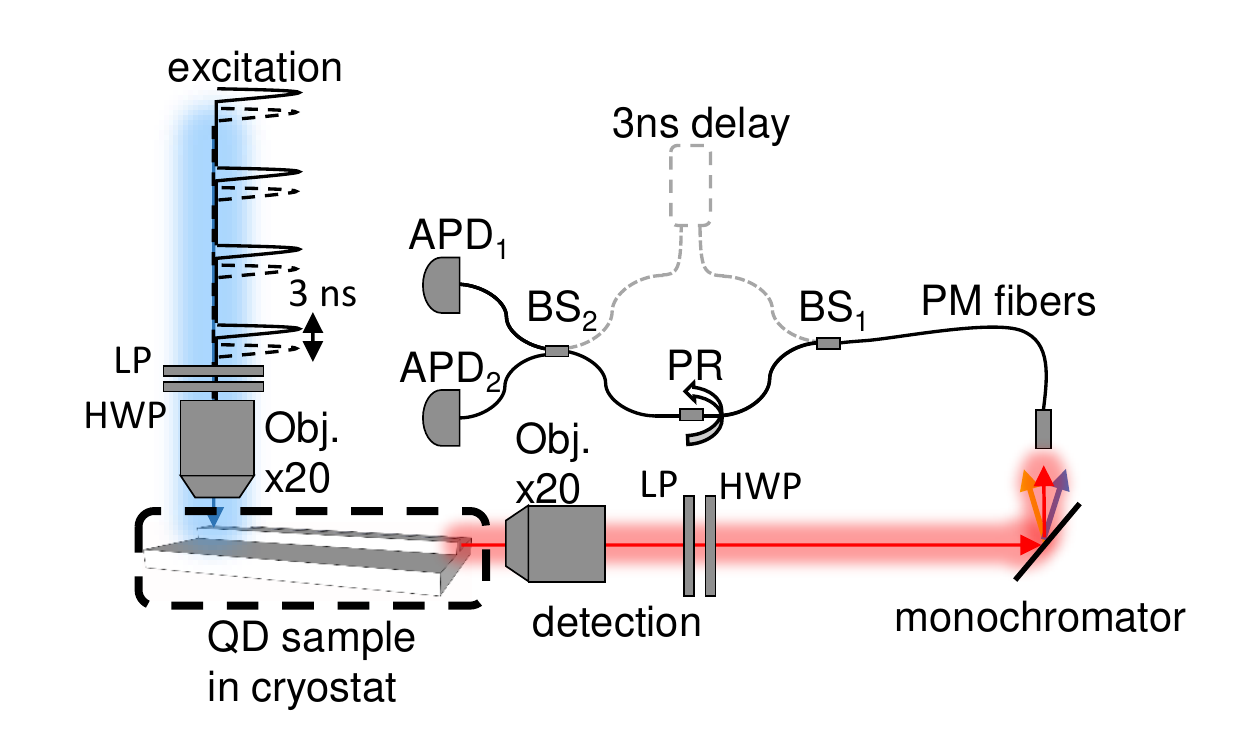}
	\caption{\label{fig:setup}Optical setup. Scheme of the experimental configuration used for top excitation (blue path) and side detection (red path) photoluminescence and resonance fluorescence measurements. In case of two-photon interference experiments, a QD was excited twice every laser pulse cycle with a delay of 3~ns, and the subsequently emitted photons, spatially and temporally overlapped in an unbalanced Mach-Zehnder interferometer (dashed lines) utilizing polarization maintaining (PM) fibers and beam-splitters (BS). For signal detection, two avalanche photo-diodes (APD) with 350~ps response time were used. For polarization control in free space, half-wave-plate (HWP) combined with a linear polarizer (LP) were used, while for polarization rotation (PR) in the fiber-based HOM interferometer ceramic sleeve connectors between two fiber facets were used allowing to align fast and slow axis at a desired angle.   
	}
\end{figure}

\textbf{Auto-correlation and two-photon interference experiment details.}
In order to characterize investigated source purity and indistinguishability the resonance fluorescence signal is first passed through a monochromator to filter out (spectral width $\sim$50~pm/70~$\mu$eV) broader laser profile and phonon sidebands, then it is coupled into a single-mode polarization maintaining fiber. Next, the light is introduced into a fiber-based unbalanced Mach-Zehnder interferometer for the two-photon interference measurements in the Hong-Ou-Mandel (HOM) configuration. To control the polarization of photons introduced into the second beam-splitter of the interferometer, a ceramic sleeve connector between two fiber facets is used allowing to align fast and slow light axis at the desired angle, and thus prepare photons in cross and parallel polarizations. For the auto-correlation measurements, one of the interferometer arms is blocked. Outputs ports are coupled to the two single-photon counting avalanche photo-detectors (APD) with a 350~ps temporal resolution. The photon correlation events are acquired by a multi-channel picosecond event timer. For the time-resolved experiments, a fast APD is used with a response time of $\sim$40~ps.

\section{Simulations details} 
\label{app:1}
\textbf{Methods.} We calculated an overall efficiency of a whole device by three-dimensional (3D) finite-difference-time-domain (FDTD) method. We used home-made FDTD software. The overall efficiency is defined as the probability to detect one photon from a QD through a waveguide into the collecting objective lens. Three efficiencies, coupling efficiency, output efficiency, collection efficiency, determine the overall efficiency. The QD emitter is modeled by a linearly polarized dipole source aligned along the transverse-electrically polarized waveguide mode. The dipole is placed at the center of the waveguide core. In the calculation, we assume a wavelength of 930 nm. The refractive indexes of GaAs and AlGaAs are set to 3.5652 and 3.0404, respectively. In order to represent infinite free space in the simulations, the uniaxial perfectly matched layer (UPML) was used as the absorbing boundary condition. On the other hand, to calculate the electric field distribution of the waveguide modes at a wavelength of 930~nm, a periodic boundary condition is used for the waveguide direction instead of the UPML boundary condition.

The coupling efficiency is defined as the ratio between the coupled photons into the waveguide mode and the emitted photons from a QD. The efficiency is estimated over the waveguide length of 40~$\mu$m by 3D FDTD. On the other hand, the output efficiency, the ratio between the escaped photons from the waveguide output port into free space and the coupled waveguide photons, is directly calculated by integrating outgoing Poynting vectors out of the waveguide. The collection efficiency, the amount of the collecting photons by the objective lens (NA = 0.4) placed in front of the output port among the escaped photons from the waveguide, is obtained by calculating far-field distribution, showing the emission directions of photons.

\textbf{QD-waveguide coupling efficiency.} 
By performing three-dimensional FDTD calculations we investigated different DBR waveguide designs for maximized QD-waveguide coupling efficiency. Figure~\ref{fig:Eff}(a) shows simulations of the QD emitted light coupling efficiency $\eta$ into both WG arms vs width of the WG with fully etched bottom DBRs. Even though coupling efficiency itself is almost independent of the WG dimensions, the ratio of particular loss channels very strongly depends on the WG height. In case of small heights most if the light leaks into sides and bottom of the WG, while in the case of fully etched WGs, losses are almost explicitly related to the bottom substrate.

\begin{figure}[t]
	\includegraphics[width=6in]{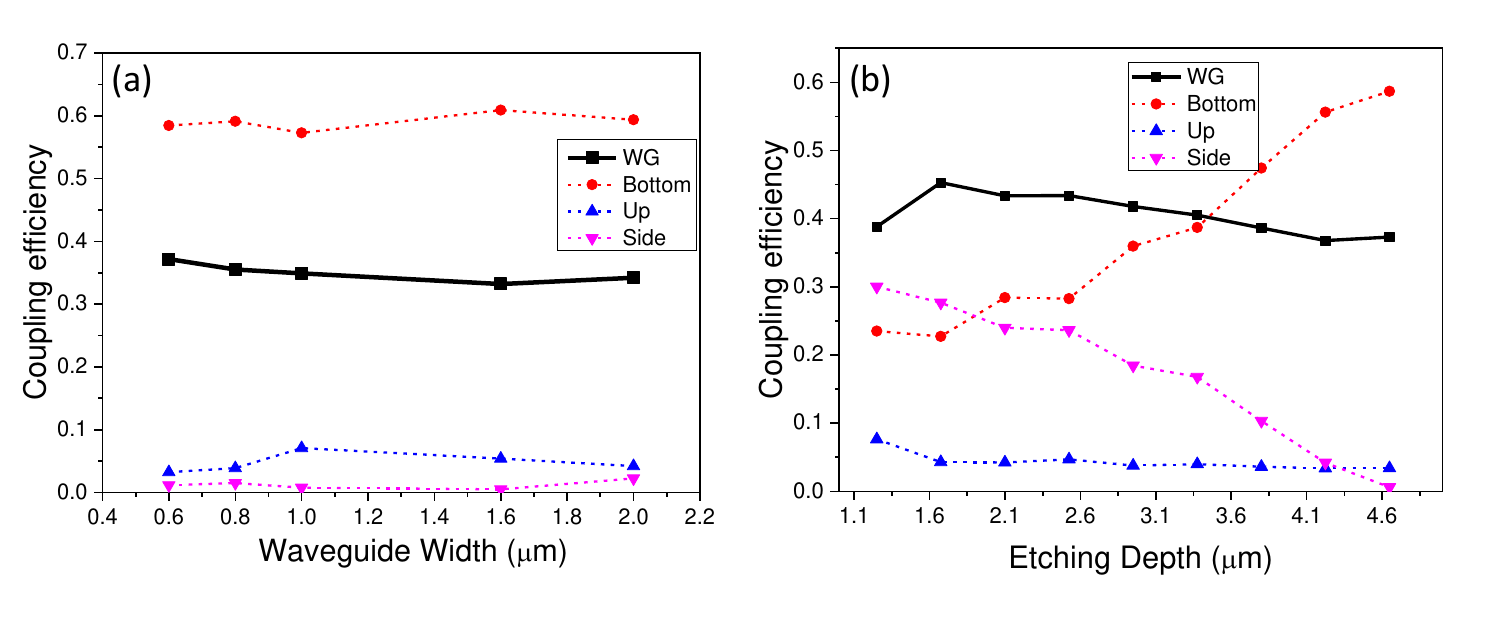}
	\caption{\label{fig:Eff}Simulations of QD-WG light coupling efficiency (into both WG arms) vs (a) width and (b) etching height for DBR WG device. Simulations have been performed for fully etched WG in case of width variation and 0.6~$\mu$m width in case of etching depth variation.}
\end{figure}
In Fig.~\ref{fig:Eff}(b) simulations of the QD emitted light coupling efficiency $\eta$ into both WG arms vs WG height for 0.6~$\mu$m WG width are shown. A flat maximum of around 40-45\% coupling efficiency can be observed for the 1.2-3.0~$\mu$m WG heights. Maximal values of around 22\% coupling efficiency into one WG arm in our case are defined mainly by the critical angle of the DBRs ($\sim$19~deg), at which the reflectivity rapidly decreases. By integration of our DBR waveguides with low-refractive-index layers coupling efficiencies could be potentially further improved.

\textbf{Single mode guiding in DBR waveguides.} In Fig.~\ref{fig:refIndex} the effective refractive index vs WG width is shown, calculated for the fundamental (TE$_1$) and first excited (TE$_2$) TE mode of the 1.25~$\mu$m height DBR waveguide at 900 and 940~nm wavelengths. As can be clearly seen, our DBR waveguides should operate in the single-mode regime for widths smaller than 0.93$\pm$0.01~$\mu$m at 900~nm wavelengths and widths smaller than 1.00$\pm$0.05~$\mu$m at a 940~nm wavelength. Taking into account simulations and fabrication inaccuracies, it should be possible to obtain single-mode operation for WGs with profile dimensions of 0.9x1.25~$\mu$m under cut-off wavelengths of 900~nm.   

\begin{figure}[h]
	\includegraphics[width=3.4in]{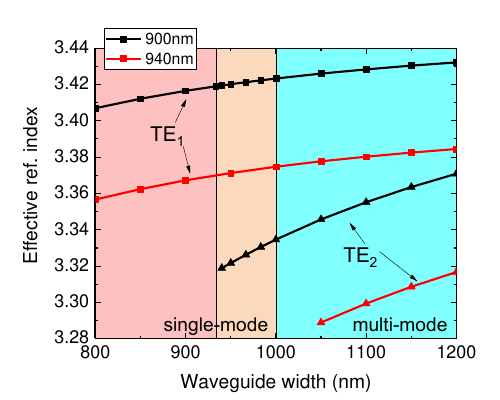}
	\caption{\label{fig:refIndex} Effective refractive index vs waveguide width calculated for the fundamental (TE$_1$) and first excited mode (TE$_2$) in DBR waveguide. At wavelength 940~nm we can obtain single-mode guiding for WG widths below 1.00$\pm$0.05~$\mu$m, while at a 900~nm wavelength for widths below 0.93$\pm$0.01~$\mu$m.  
	}
\end{figure}

\textbf{Total output extraction efficiency.} In order to check the properties of single photons generated on-chip, the fluorescence signal has to be out-coupled from the waveguide and introduced into the optical setup. We estimated that less than 5\% (10\%) of light initially coupled to the 2.0~$\mu$m (0.8~$\mu$m) wide MM (SM) WG is effectively collected by our microscope objective, which corresponds to the (i) waveguide transmission losses ($\sim$29\% loss for 1~mm travel), (ii) WG facet-air interface loss due to the reflection $\sim$81\% ($\sim$35\%) and scattering $\sim$1\% ($\sim$15\%), and finally (iii) mode mismatch between the WG and collection optics $\sim$51\% (60\%) loss. In total, around 0.88\% (2.7\%) of photons emitted from QD$_1$ (QD$_2$) should be collected into the first lens of the optical setup.  

In this work, we focus on generating single photons with high mutual indistinguishability. It means that we cannot use all the photons generated by a QD since part of them are actually scattered by phonons making them distinguishable. This part of the QD emission spectrum (phonon sidebands) has to be removed, which forces us to use a spectral filter. We employ off-chip filtering with $\sim$100~$\mu$eV width, which based on J. Iles-Smith et al. [Nat. Photonics 11, 521 (2017)] allows to achieve $\sim$99\% indistinguishability and limits maximal obtainable efficiency to $\sim$83\%.          

\textbf{Fraction of light coupled to each mode for 1.25~x~2.0~$\mu$m MM waveguide.} In the case of the DBR waveguide with 1.25~x~2.0~$\mu$m profile dimensions, six modes are allowed for the guiding at 930 nm wavelength. Based on the FDTD calculations, if the central position of the QD would be assumed, most of the light coupled into the WG will be distributed between fundamental 1st mode (80.3\%), 3rd mode (17.2\%) and 5th mode (2.43\%), while the 2nd, 4th and 6th will have negligible coupling ratio, due to the low overlap between centrally positioned QD dipole and WG mode distribution (0.0025\%, 0.01\%, 0.0004\% respectively). 

\textbf{Purcell enhancement factor in DBR waveguides.} In order to estimate the potential Purcell enhancement factor in case of our DBR waveguides, we performed FDTD simulations by putting a dipole emitter in the centre of our DBR waveguide and calculated the output power into the WG and compared it with the output power of an emitter in bulk. The spontaneous emission rate enhancement of the emitter in the WG in respect to the bulk is 0.94 and 0.96 for 1.25~x~2.0~$\mu$m and 1.25~x~0.8~$\mu$m WG profile dimensions, respectively.

\section{Detail description of the two-photon interference histogram peaks} 
\label{app:2}
Peaks 1 and 5 of Fig.4(b) in the main text describe the process, where the first emitted photon is traveling through a shorter arm of the interferometer and second through longer one, which generates additional 3~ns delay and gives rise to bunching at $\pm$6~ns. Peaks labeled as 2 and 4 describes a process of the first and second photon traveling through the same arm in the interferometer, which is twice more probable than an earlier process and gives rise to the bunching at $\pm$3~ns. Peak 3 which describes the situation in which first emitted photon is delayed in the longer arm, and the second one travels through the shorter arm so that both photons superimpose into the second beam splitter. According to the above, for perfectly indistinguishable single photons the peak area ratio would be 1:2:0:2:1, in contrary to perfectly distinguishable single photons 1:2:2:2:1 ratio. The non-central peak clusters follow Poissonian statistics with the 1:2:6:2:1 ratio. In order to quantitatively evaluate the peak 3 area in respect to the neighboring peaks experimental data have been fitted with two-side exponential decay functions with a fixed area ratio of 1:3:6:3:1 and 1:2:X:2:1, for side and central peak clusters, respectively (red solid line). The zero delay peak relative area - X, and the Poissonian level were fitting parameter, while the decay time constant was fixed to value obtained in the time-resolved measurements. It needs to be noted here, that due to the 3~ns delay between the pulses, peaks positioned at $\pm$6~ns are imposed with 6~ns delayed peaks coming from the neighboring 12.2~ns delayed pulses, so that central cluster 2:2:0:2:2 ratio rather than 1:2:0:2:1 is visible. In this context inclusion of neighboring peaks into histogram analysis is needed.  

\section{Two-photon interference histogram fitting and visibility extraction} 
\label{app:3}
In order to extract peaks areas of the two-photon interference histogram, experimental data have been fitted with the two-sided exponential decay function with a central peak intensity ratio of 1:2:X:2:1, where $X$ corresponds to the zero delay peak area, and 1:4:6:4:1 ratio for non-central peaks

\begin{equation}
\begin{split}
F(t) = &\frac{A}{2} \left(X e^{\frac{-\left|t\right|}{\tau_{dec}}}
+ 2 e^{\frac{-\left|t\pm\tau_{pd}\right|}{\tau_{dec}}}
+ e^{\frac{-\left|t\pm 2\tau_{pd}\right|}{\tau_{dec}}} \right) \\
+ \sum_k &\frac{A}{2} \left( 6 e^{\frac{-\left|t\pm k\tau_{rr}\right|}{\tau_{dec}}}  
+ 4 e^{\frac{-\left|t\pm k\tau_{rr} \pm \tau_{pd} \right|}{\tau_{dec}}} 
+ e^{\frac{-\left|t\pm k\tau_{rr} \pm 2\tau_{pd} \right|}{\tau_{dec}}} \right),
\end{split}
\end{equation}
where $A$ is corresponding to number 2 (4) central cluster peak amplitude of intensity, $\tau_{dec}$ is a resonance fluorescence decay time, $\tau_{pd}$ is a delay time between the pulse trains (3~ns) and $\tau_{rr}$ is a repetition time of the laser (12.2~ns). The first line of the equation corresponds to the central cluster of peaks, while the second is the sum over non-central clusters. For the fitting procedure all values beside the $X$ and $A$ (fitting parameters), were fixed. The visibility has been calculated based on the central peak areas at zero delay $A_3$ and $\pm$3~ns delay $A_2$ and $A_4$, which have been directly obtained from the fitting parameter $X$
\begin{equation}
V_{exp} = 1-\frac{2A_3}{A_2 + A_4} = 1 - \frac{X}{2}.
\end{equation}
To take into account influence HOM interferometer imperfections limiting maximal obtainable experimentally visibilities, we corrected $V_{exp}$ values according to
\begin{equation}
V = V_{exp}\cdot\frac{R/T+(R/T)^{-1}}{2(1-\varepsilon)^2},
\end{equation}
where $R/T$ is beam-splitter reflectivity/transmission ratio and $(1-\varepsilon)$ is contrast of the HOM interferometer. The uncertainty of the $V$ value is calculated based on the $X$ fitting precision. We note that no corrections for the $g^{(2)}(0)$ value nor the background were implemented.

\begin{table*}[h]
	\caption{\label{tab:table}Summarized values of recorded single photon purity and indistinguishability.}
	\begin{ruledtabular}
		\begin{tabular}{cccccccc}
			Waveguide&Emitter&Energy&$R/T$&$1-\varepsilon$&$V_{exp}$&$V$&$g^{(2)}(0)$\\ \hline
			MM&QD$_1$&1.3169~eV&1.12&0.9897&0.949&0.975$\pm$0.005&0.009$\pm$0.002 \\
			SM&QD$_2$&1.3206~eV&1.14&0.9915&0.923&0.950$\pm$0.005&0.04$\pm$0.005
		\end{tabular}
	\end{ruledtabular}
\end{table*}

\section{Laser contribution into resonance fluorescence spectra} 
\label{app:5}

The excitation laser background intensity was tracked by observing a broader laser profile in resonance fluorescence spectra (Fig.~\ref{fig:RFlaser}) which is clearly visible in linear-log-scale (insets of Fig.~\ref{fig:RFlaser}). In case of QD$_1$ much better signal-to-laser-background ratio was achieved for $\pi$-pulse pumping than for QD$_2$, which might be related to the QD emission energy located closer to the vertical cavity resonance (1.316~eV). 

\begin{figure}[h]
	\includegraphics[width=6in]{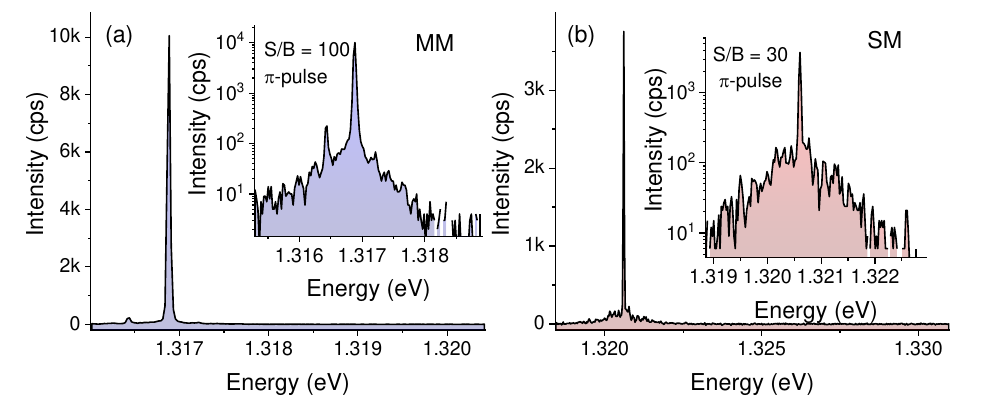}
	\caption{\label{fig:RFlaser}Resonance fluorescence spectra for (a) QD$_1$ coupled to MM waveguide and (b) QD$_2$ coupled to SM waveguide. Insets: Emission spectra in a linear-log scale with visible much broader excitation laser profile.  
	}
\end{figure}

\begin{figure}[h]
	\includegraphics[width=5in]{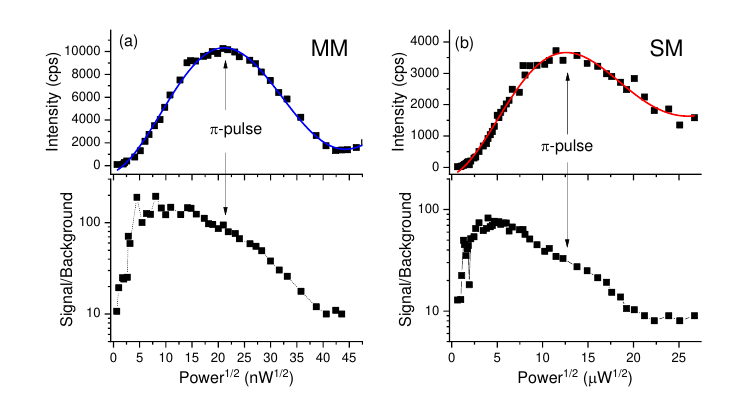}
	\caption{\label{fig:RFpower}Resonance fluorescence intensity and signal-to-background ratio in function of square root of excitation power for (a) QD$_1$ coupled to MM waveguide and (b) QD$_2$ coupled to SM waveguide.  
	}
\end{figure}

By utilizing both the intrinsic and polarization filtering we were able to obtain a signal-to-background (S/B) ratio in the range of 10-150 and 8-80 for QD$_1$ and QD$_2$, respectively, depending on the excitation power. In Figure \ref{fig:RFpower}(a) and (b) the resonance fluorescence peak intensity versus the square root of the incident power are shown. Based on the resonance fluorescence spectra, the signal-to-background ratio was estimated for a given set of powers and plotted in Fig.~\ref{fig:RFpower}(c) and (d). The best S/B ratios were obtained for powers corresponding to the $\pi$/2-pulse. Under $\pi$-pulse excitation S/B ratio of around 100 and 30 for QD$_1$ and QD$_2$ have been obtained, respectively. We note here, that S/N ratio might to large extend limit single photon purity and indistinguishability values observable in HBT and HOM experiments.

\section{Temperature tuning of the QD emission} 
\label{app:4}
PL spectra for QD$_1$ and QD$_2$ exhibit intensity modulations, with a constant period of 200~$\mu$eV and 370~$\mu$eV corresponding to Fabry-Perot (FP) type oscillations between GaAs-air interfaces localized on the two ends of the waveguides. By tuning the QD emission energy with temperature along the FP resonances significant changes of the PL intensity can be observed as visible in Figures~\ref{fig:temp}(a) and (b).

\begin{figure}[h]
	\includegraphics[width=6in]{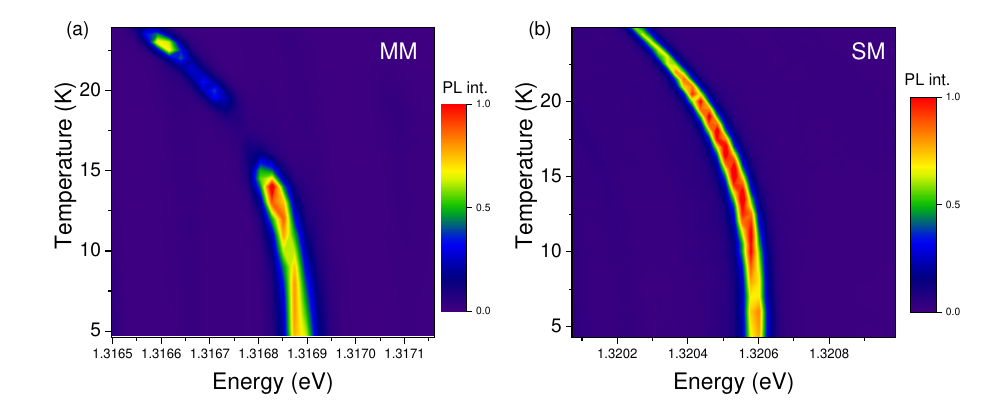}
	\caption{\label{fig:temp}Non-resonant 2D PL intensity vs temperature spectra (colorbar in linear scale) for (a) QD$_1$ coupled to MM waveguide and (b) QD$_2$ coupled to SM waveguide, showing intensity modulation related to the Fabry-Perot type cavity.  
	}
\end{figure}

\end{widetext}

\end{document}